\documentclass{article}[10pt]
\usepackage{graphics,graphicx,epsfig,color}
\usepackage{a4wide}
\usepackage{geompsfi-pdf}
\usepackage{amsmath,amsfonts}
\usepackage{mathrsfs}
\usepackage{float}
\include{today}
\long\def\drop#1{}

\begin{document} 
\renewcommand{\thefootnote}{\fnsymbol{footnote}}
\title{Multilayered folding with voids}
\author{Tim Dodwell\footnotemark[1], Giles Hunt\footnotemark[1], Mark Peletier\footnotemark[2] \footnotemark[1] and Chris Budd\footnotemark[1]}
\footnotetext[1]{Centre for Nonlinear Mechanics, University of Bath, Bath BA2 7AY, UK.}
\footnotetext[2]{Institute of Complex Molecular Systems and Department of Mathematics and Computer Science, Technische Universiteit Eindhoven, PO Box 513, 5600MB Eindhoven, The Netherlands.}
\date {\today}

\label{firstpage} 
\maketitle 
\begin{abstract} 
In the deformation of layered materials such as geological strata, or stacks of paper, mechanical properties compete with the geometry of layering. Smooth, rounded corners lead to voids between the layers, while close packing of the layers results in geometrically-induced curvature singularities. When voids are  penalized by external pressure, the system is forced to trade off these competing effects, leading to sometimes striking periodic patterns. 

In this paper we construct a simple model of geometrically nonlinear multi-layered structures under axial loading and pressure confinement, with non-interpenetration conditions separating the layers.  Energy  minimizers are characterized as solutions of a set of fourth-order nonlinear differential equations with contact-force Lagrange multipliers, or equivalently of a fourth-order free-boundary problem. We numerically investigate the solutions of this free boundary problem, and compare them with the periodic  solutions observed experimentally.

\begin{center}
\end{center}
\end{abstract}
\section{Introduction}
\subsection{Folds in multilayer structures: rocks and paper}
\begin{figure}[!ht] 
   \centering
   \includegraphics[width=2.4in]{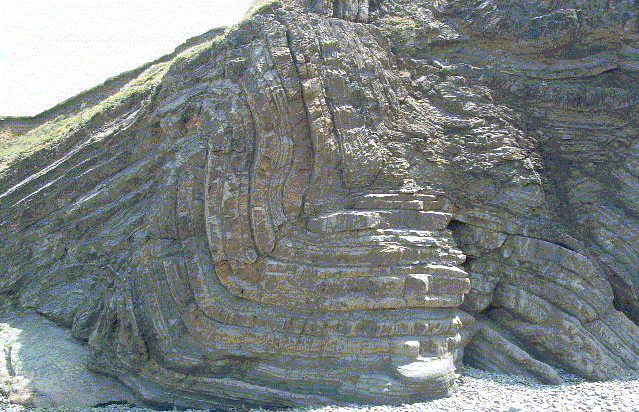}
    \includegraphics[width=2.4in]{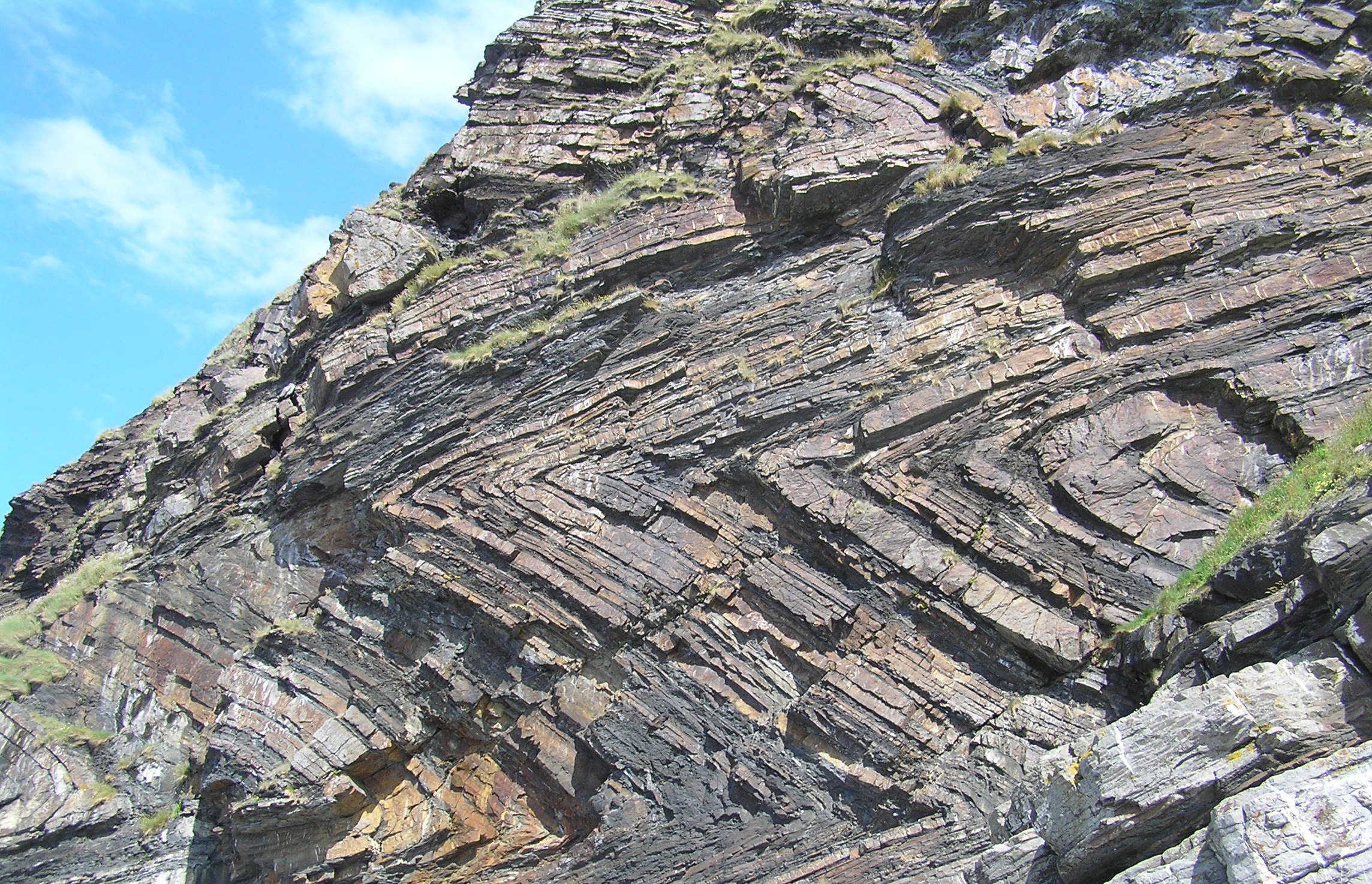}
         \caption{Folding patterns in thinly bedded shales and sandstones in Cornwall, UK. Left: parallel folding at Bude. Right: chevron folding at Millook Haven.}
     \label{fig:millook}
\end{figure}

Geological folds arise from the in-plane compression of layers of rocks from tectonic plate movement, and provide one of the overarching themes of this special issue (see also \cite{Hobbs2011,Schmalholz2011}).  Geometric constraints following the need for layers to fit together generate a number of distinctive forms of folding, not seen in single-layered systems (see for example Price \& Cosgrove \cite{PriceCosgrove90}). A couple of illustrative examples are shown in Fig.~\ref{fig:millook}. 
On the left is an example of {\em parallel folding} where, as the eye moves from bottom left to top right, the curvature of the layers increases until eventually it becomes too tight, and forms a distinctive \emph{swallowtail singularity}~\cite{levelsets}, after which the layers are seen to crumble.  On the right, the geometric constraints have been overcome by the rocks adopting straight limbs and sharp corners, in what is known as {\em chevron folding} \cite{voids1}. This  phenomenon is strongly reminiscent of {\em kink banding}, seen also in other contributions to this special issue \cite{Pinho2011,Wadee2011}. 

\begin{figure}[!ht] 
   \centering
  \includegraphics[width=4in]{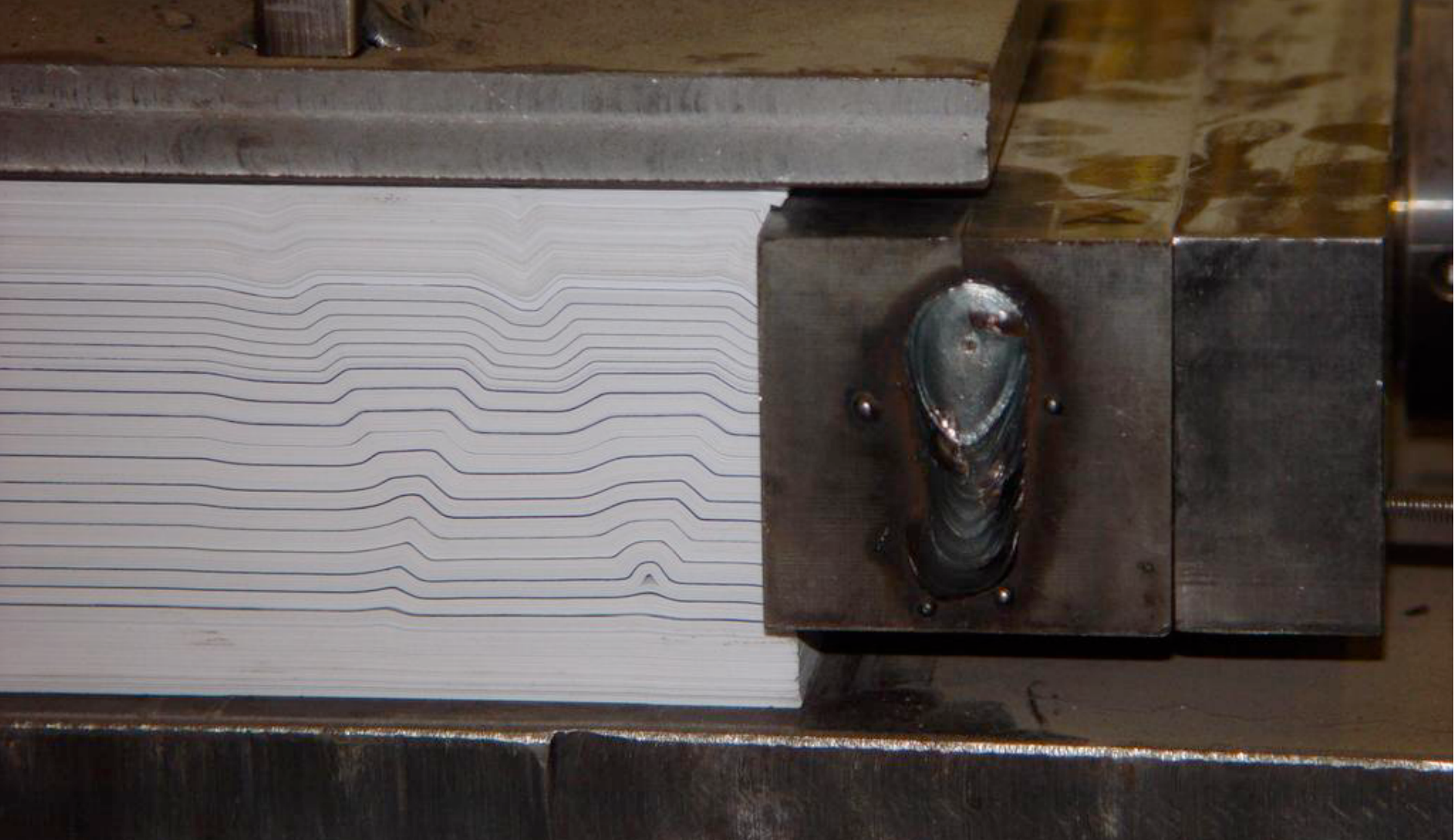}  
  \includegraphics[width=4.5in]{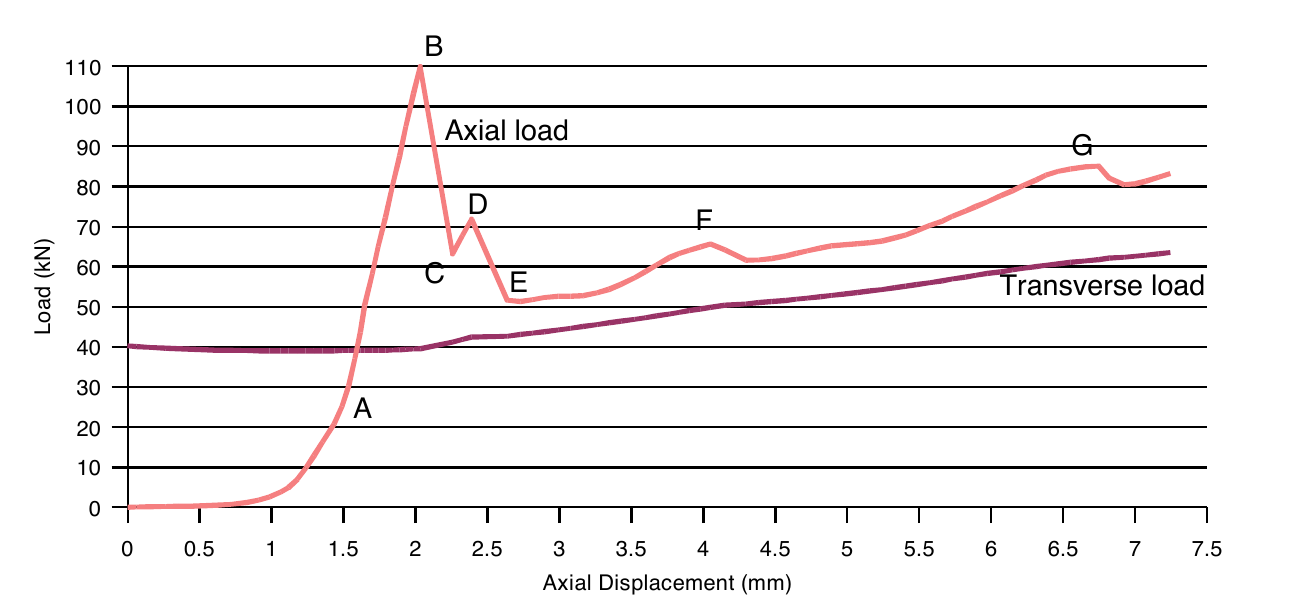} 
     \caption{Top: Kink banding in compressed layers of paper. Deformation is forced in about 625 sheets of paper from a sample of 800. Bottom: output diagram: load \emph{vs.} end displacement. }
     \label{fig:kinks}
\end{figure}

The sequence of events leading to such structures is of course lost in geological time, but small-scale experiments on layered materials such as paper do provide some suggestions. The outcome of one such experiment \cite{kinks3} is shown in Fig.~\ref{fig:kinks}.
About 800 half sheets of A4 cut longitudinally were squashed together under a {\em transverse load} between steel plates, and then slowly loaded in the longitudinal direction by a second applied {\em axial load}. Load cells were used to record both loads at one-second intervals, and an in-plane transducer registered the axial end displacement. To aid visualization, a black-edged sheet was introduced approximately every 25 layers.

\begin{figure}[!ht]
\vspace{0.5cm}
\centering
\quad
  \includegraphics[width=2.2in]{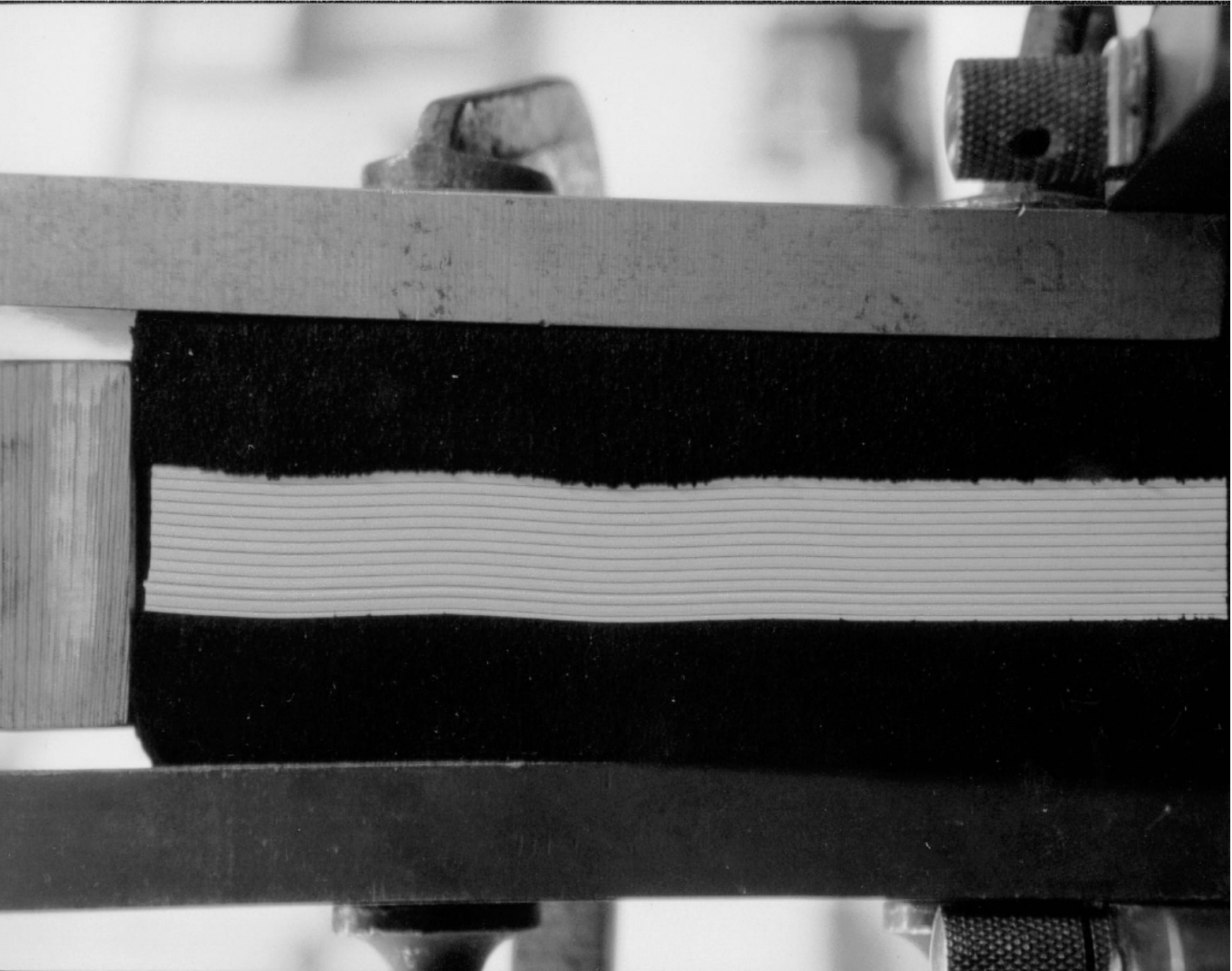} \hspace{1cm}
   \includegraphics[width=2.2in]{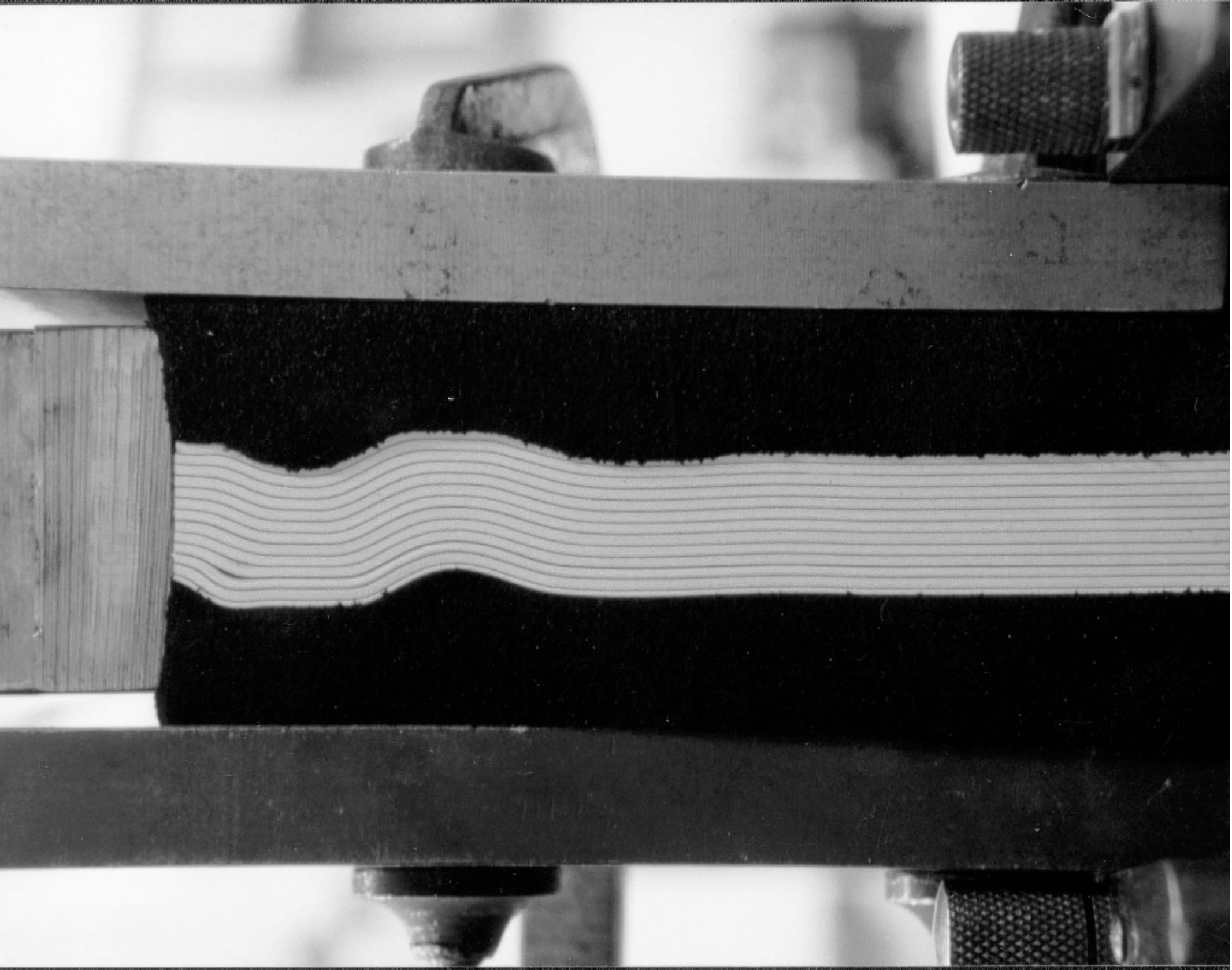}  \\[3mm]
   \includegraphics[width=2.2in]{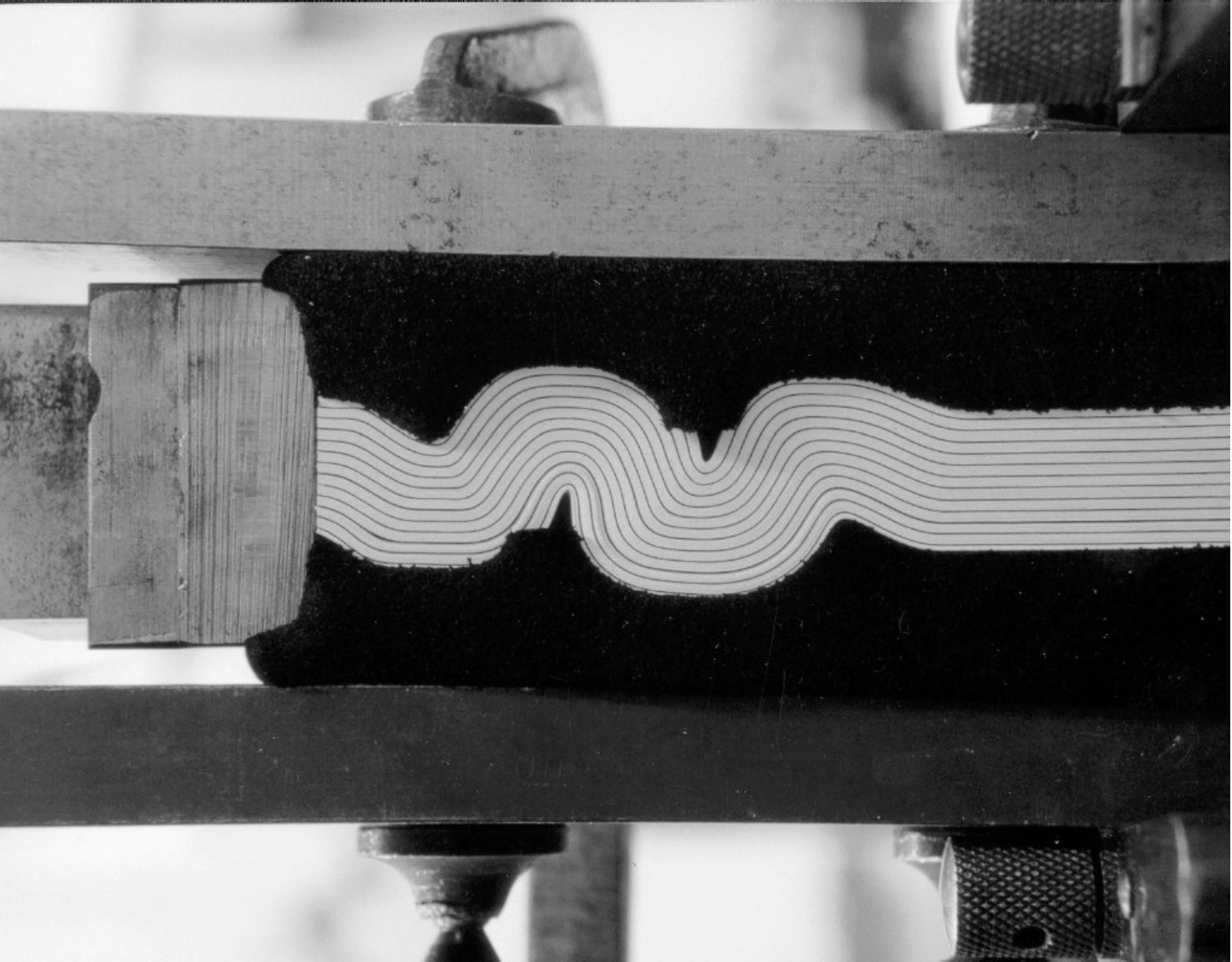} \hspace{1cm}
  \includegraphics[width=2.2in]{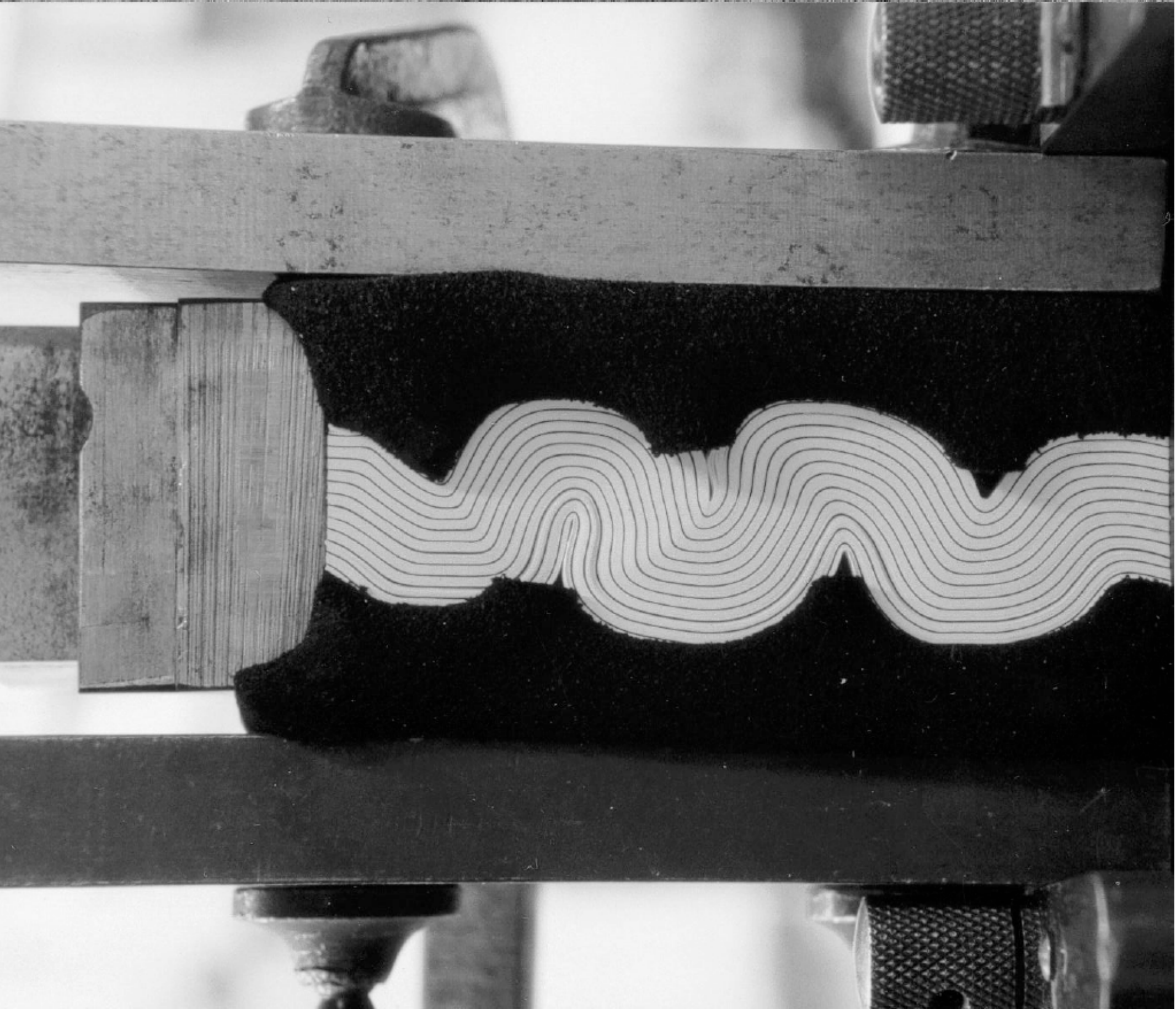}  
\vspace{5mm}
\caption{A compression test with a weaker, but stiffening, foundation (foam). Observe the parallel folding and and the serial or sequential buckling.}
\label{fig:serial}
\end{figure}

Output from a typical experiment is shown at the bottom. Kink bands formed in sequence from the loaded end, starting at the point B, followed in turn by D, F and G, seen on the output diagram. Each band forms with a clear-cut instability, observed by an corresponding instantaneous fall in applied axial load and followed by subsequent re-stabilization or \emph{lock-up}.
By comparison, Fig.~\ref{fig:serial}
shows various stages in the loading history of a similar experiment, but with soft foam layers replacing the upper and lower (stiff) supporting foundation, such that the overburden pressure and accompanying stiffness are much reduced. The folds still develop and lock-up sequentially, but the sharp corners and straight limbs of the kink-bands are replaced by curved layers, as in the parallel folding example of Fig.~\ref{fig:millook}; lock-up is here associated with the tightening of the curvature on the inside of a fold \cite{levelsets}.  

\begin{figure}[ht] 
    \centering
     \includegraphics[width=1.5in]{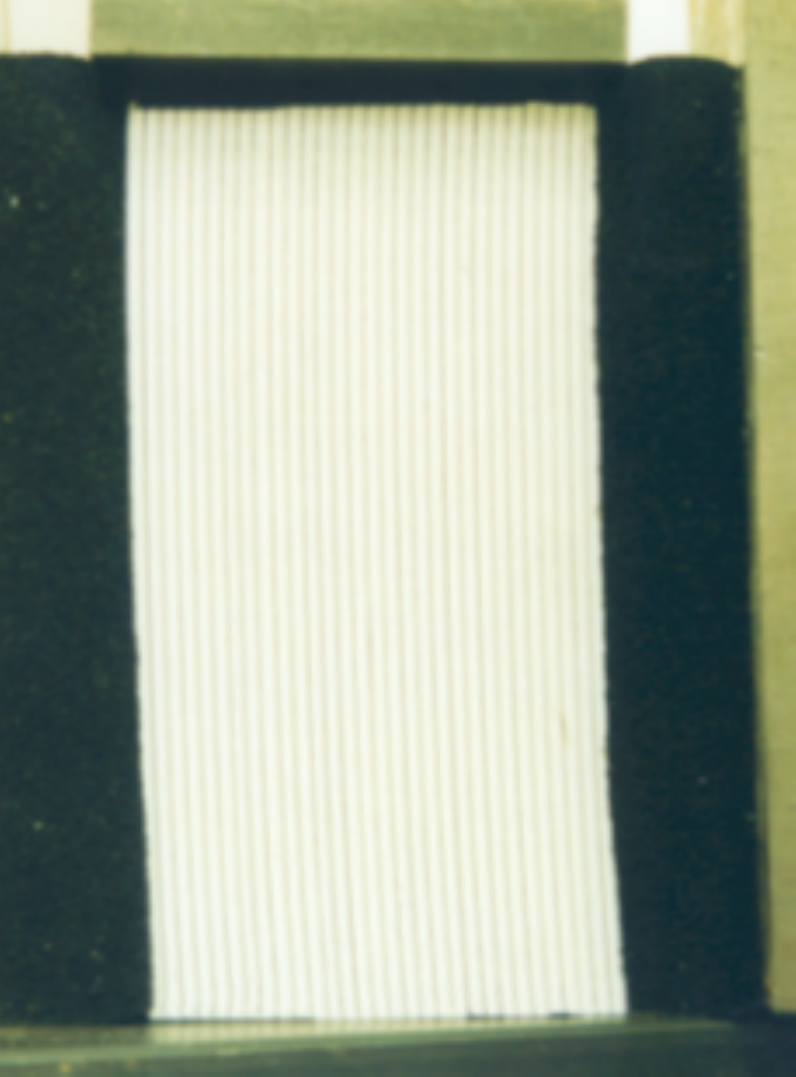} 
     \hspace{0.5cm}
     \includegraphics[width=1.35in]{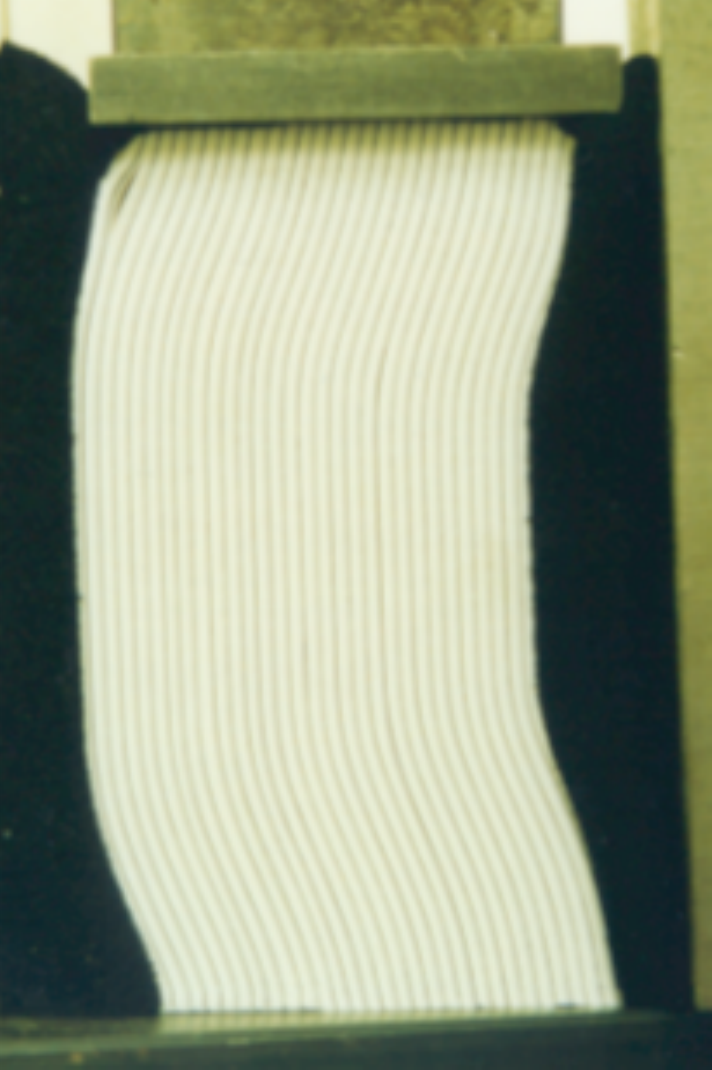}     
      \hspace{0.5cm}
     \includegraphics[width=1.35in]{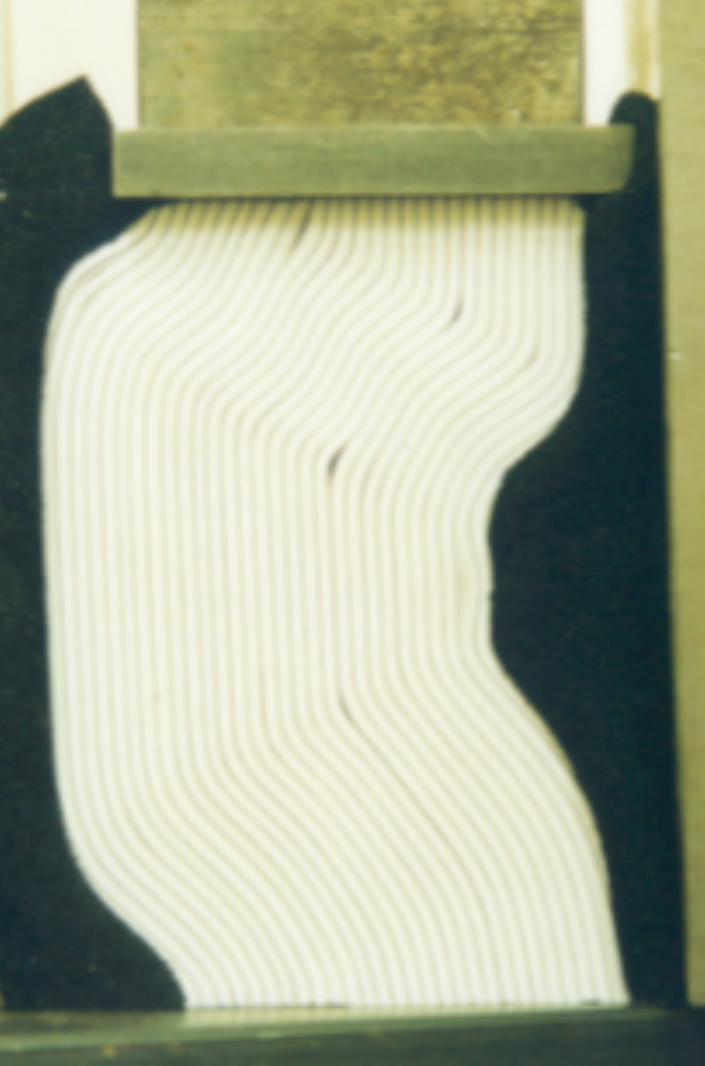}   
     \vskip1em 
  \includegraphics[width=4in]{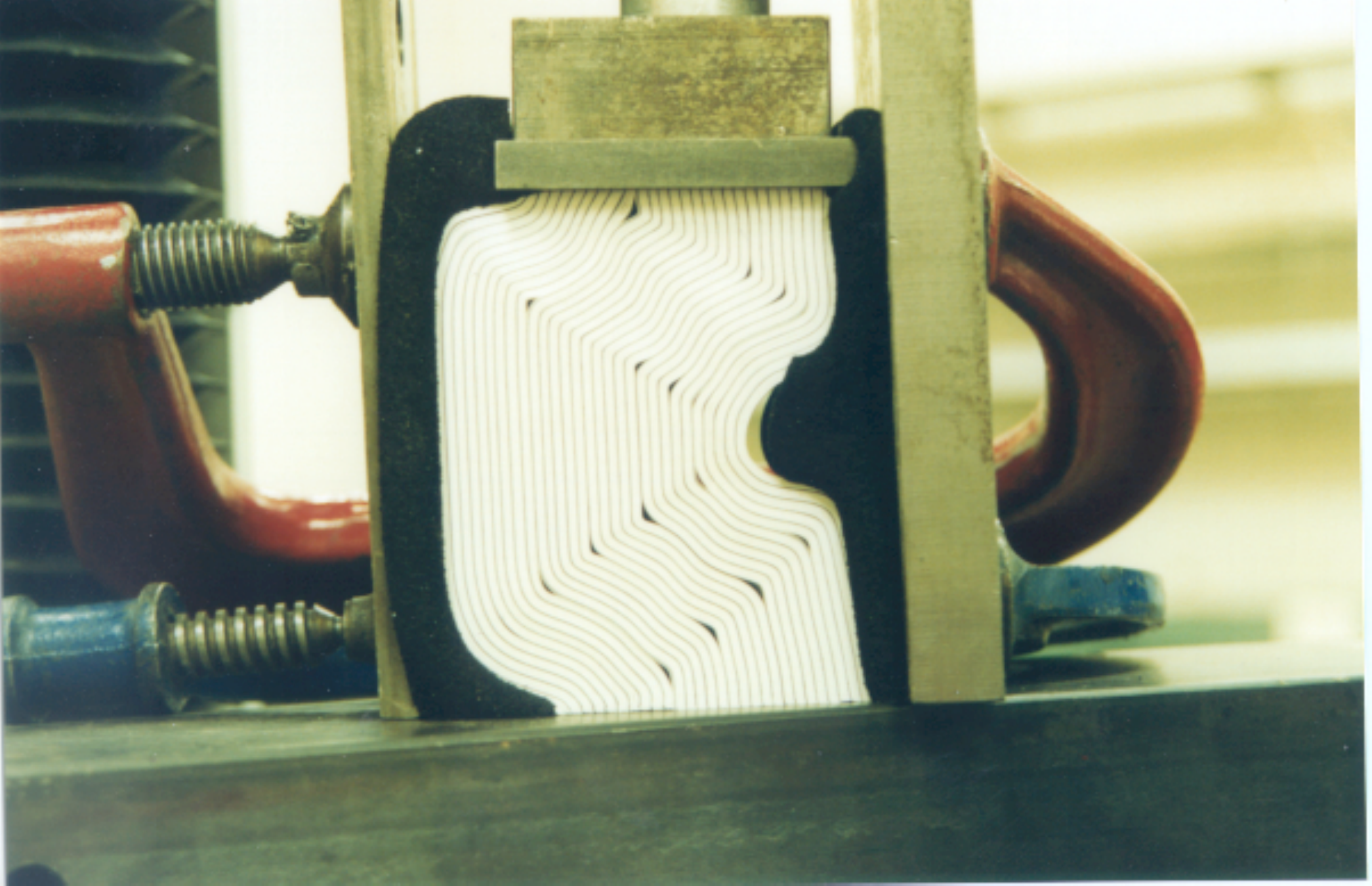}  
     \caption{A compression test with low initial overburden pressure. Note the periodically positioned voids at kink band corners in the final picture.}
      \label{fig:test1-3}
     \label{fig:test5}
\end{figure}

The test shown sequentially in Fig.~\ref{fig:test1-3}
starts under conditions similar to parallel folding, but evolves to a situation more like kink-banding.  Foam is again  used as a supporting medium, but at the onset of loading only a small overburden pressure is applied, so effectively an Euler buckle initially appears over one half wave; this is just apparent in the left-hand image of Fig.~\ref{fig:test1-3}. The long wavelength means that deflections into the foundation grow relatively rapidly, and quickly bring the outer steel plates into play, stiffening the foundation. This strongly nonlinear effect both flattens the crest of the wave, and instigates two outlying kink bands orientated across the sample at about $30^\circ$ to the transverse direction, seen most clearly in Fig.~\ref{fig:test5}. Although these appear similar to those of Fig.~\ref{fig:kinks}, they are formed under quite different circumstances, as follows. 

When a kink band is orientated obliquely across the sample, as it displaces from its undeflected (flat) state, the layers initially dilate within the band \cite{kinks3}. Those seen in Fig.~\ref{fig:kinks} have used this property to select their angles of orientation across the sample, such that maximum available dilation exactly cancels the pre-compression induced by the overburden pressure and friction is eliminated. They propagate across the sample in this optimum configuration, in which work done against friction is minimized; this orientation is then maintained as the band develops and eventually locks-up.

The same is not true for the system of Fig.~\ref{fig:test5}. The development of the band proceeds slowly from the onset of loading, and its orientation is not freely chosen but imposed from the start, and dictated by factors such as sample length and overall thickness. If the angle of orientation of the band is greater than its optimum value, voids will result as seen here. Note that, although the introduction of a black-edged sheet clearly helps to induce voids, they appear as a periodic sequence every five black sheets, or about once every 125 layers. Very similar effects are also found in paperboard creasing and folding, as seen in Fig.~8 of Beex \& Peerlings~\cite{BeexPeer} (this special issue). 

\subsection{A model of multilayer folding}

In this paper we focus on how bending stiffness, overburden pressure, and geometric constraints combine to cause voids, or not. The intriguing periodicity of the voids in Fig.~\ref{fig:test1-3} serves as an example: is it possible to explain, or even predict, the way voids concentrate at certain layers? 

We do so by formulating a simple model of a multilayer material in which we eliminate all but the essential aspects. The layers are described by (linear-elastic) Euler-Bernouilli beams, which are inextensible and laterally incompressible; the contact condition is `hard', and the overburden pressure is modeled by a simple global energetic expression (see Section~\ref{sec:multi} below). Geometric nonlinearity is preserved, and the system will undergo large rigid-body rotations.

Such a model is highly simplified in comparison to other studies in the literature (see~\cite{HudlestonTreagus10} for a recent review), and differs from a large part of the literature by assuming elasticity rather than viscosity. 
By studying such a simplified model, however, the appearance and size of voids are precisely defined, and we have detailed access to their properties.
As for the elasticity, we believe that for the type of question that we pose here elastic and viscous materials behave sufficiently similarly on the relevant time scales. In other work we have seen how the assumption of elasticity allows for a precise analysis~\cite{HuntPeletierWadee00,kinks3,levelsets,parallel}.

Our previous study of the behaviour of voids concentrated on a single infinite elastic inextensible beam forced into a V-shaped corner under uniform pressure $q$~\cite{voids1}. We generalize this model to multilayer folding in two steps.  First we recast the  model of voiding of~\cite{voids1} to embrace finite length beams and axial loads. The introduction of axial load negates the favourable properties found for minimizers of the purely pressured case: minimizers are no longer necessarily convex, symmetric, or have a single void space. It is shown that, under various loading conditions, minimum energy can lie with buckling on a local length scale \cite{fold,HMW}, demonstrating higher order \emph{accommodation structures}. 

\begin{figure}[!h]
   \centering
  \includegraphics[width=2.5in]{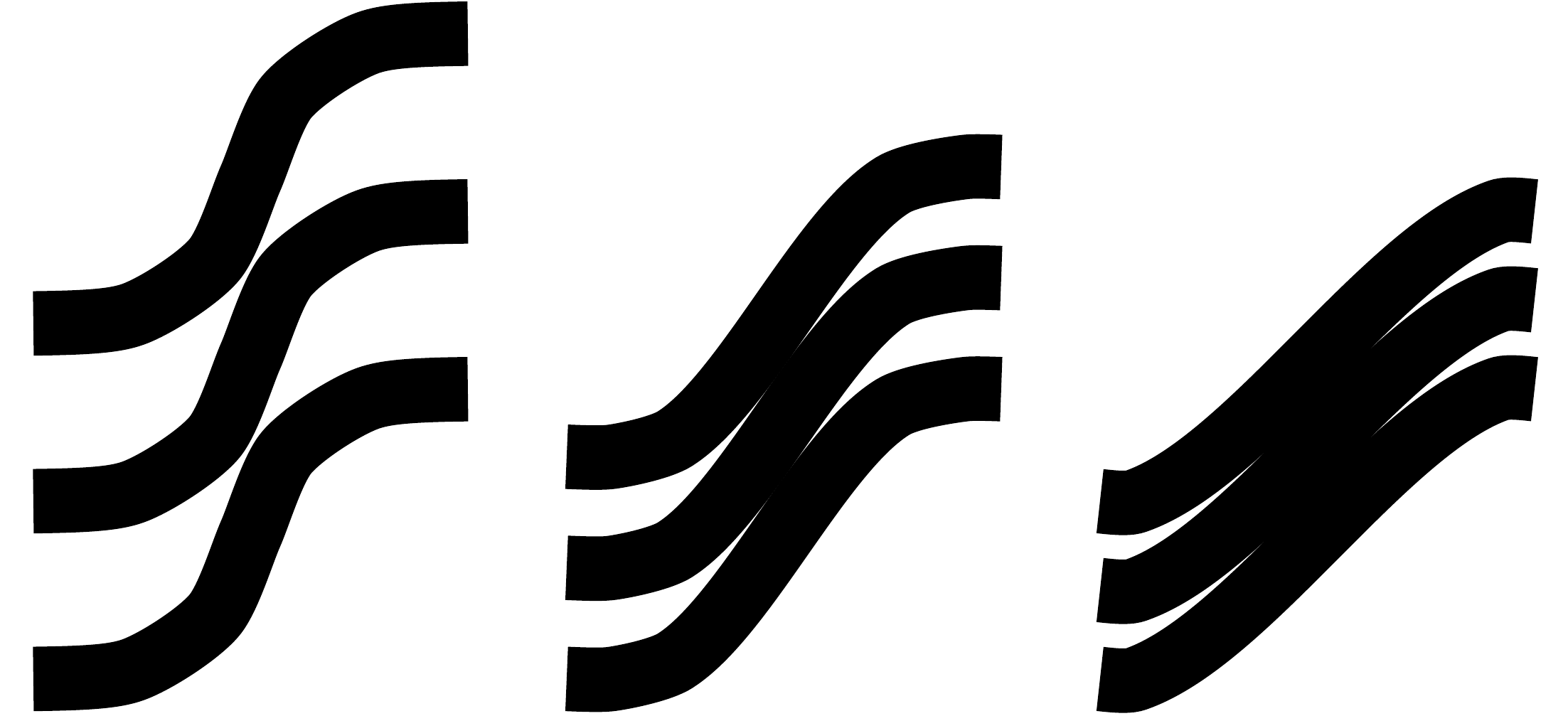}
     \caption{Sharp-angle, straight-limb folds give rise to smaller void space than rounded folds.}
   \label{fig:chevron}
\end{figure}

The main section of this paper then combines the multilayered set up presented in~\cite{levelsets} with the elasticity properties of Section~\ref{sec:singlemodel}, to capture the periodic behaviour of Fig.~\ref{fig:test5}. A first experiment in this direction is to stack identical copies of elastic layers; an elementary geometric argument suggests that reduction of the total void space could force such a stack into a straight-limbed, sharp-cornered configuration, as seen in Fig.~\ref{fig:chevron}.

In Section~\ref{sec:multi} we construct this experiment mathematically by deriving a potential energy function for a system of $N$ identical layers that are forced into a V-shaped corner by a uniform axial load $P$ and overburden pressure $q$  (see Fig.~\ref{fig:multilayereddeformation} below), to produce a so called $1$-period solution. The energy for this specific case maybe simply expressed in terms of a single layer, which leads to a similar analysis as presented in Section~\ref{sec:singlemodel}. The results match our initial elementary argument and show that the total void space reduces as the overburden pressure increases, giving deformation with straight limbs and sharp corners. If we then consider $m$-period solutions, where one void forms every $m$ layers and then repeats, we are led to an interesting question. Which $m$-periodic solution has minimal energy? Constructing an energy for $m$-periodic solutions and the subsequent variational analysis is non-trival, and we therefore leave this to a future publication. We do, however, offer some preliminary energy calculations, which suggest that the energy typically is minimized at an $m$ that is larger than one but still finite. When we choose reasonable values for the parameters in the experiments described in Fig.~\ref{fig:test5}, the predicted  value of $m$ is of the same order of magnitude as the observed value, giving us the hope there is some more interesting research to follow.
 
\section{Single layer bending and buckling}\label{sec:singlemodel}

\begin{figure}[!h] 
   \centering
   \quad
  \psfig{figure=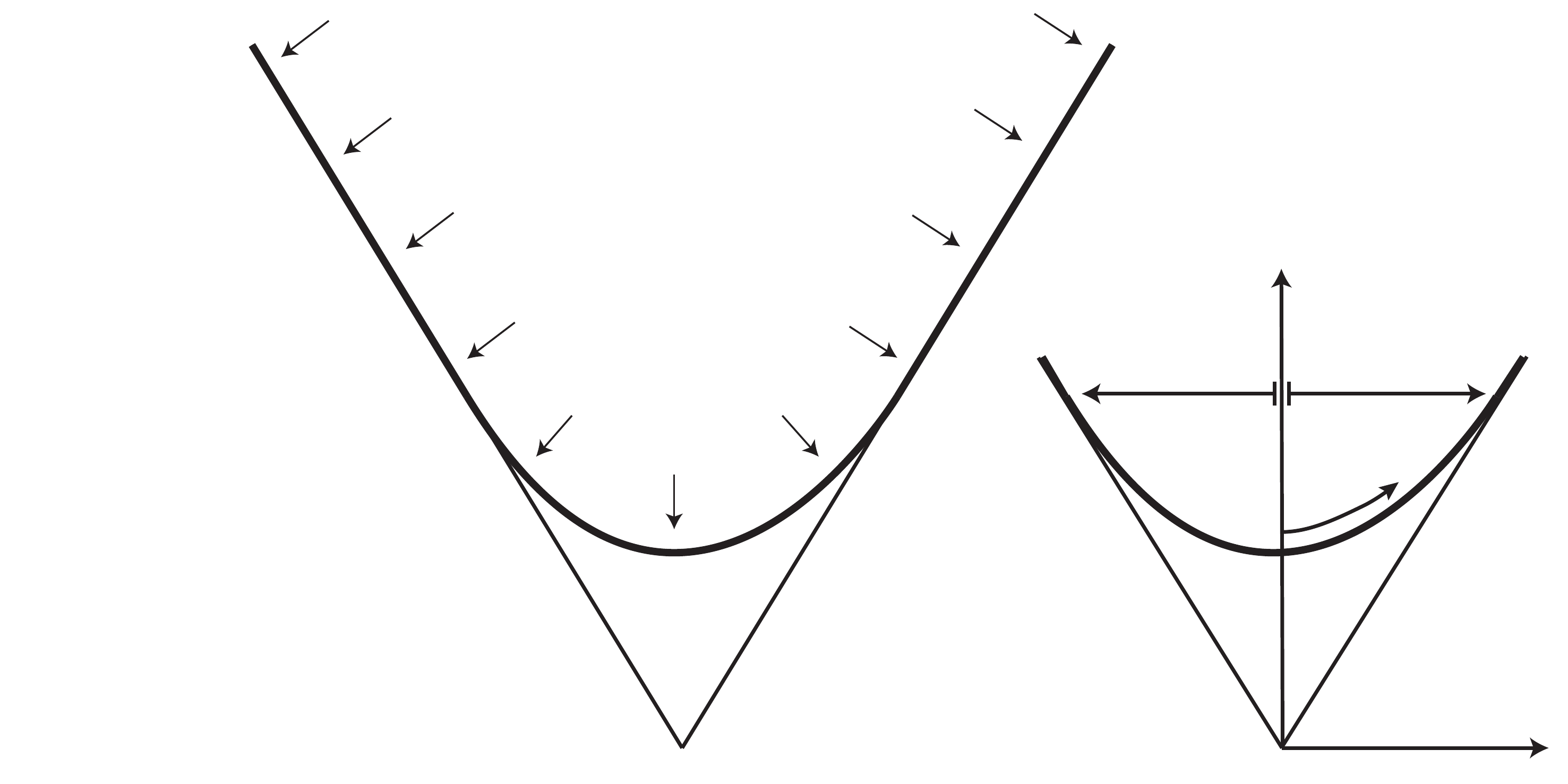,width=4in}  
     \caption{Single elastic layer, of infinite length, being forced into a corner by the action of pressure~$q$.}
     \label{fig:modelwithq}
\end{figure}

Dodwell {\em et al.} \cite{voids1} treats the system of Fig.~\ref{fig:modelwithq} as an obstacle problem with a free boundary, where the potential energy for the system is minimised subject to the pointwise obstacle constraint $w \geq f$.
The potential energy functional, $V(w)$, comprises strain energy stored in bending, $U_B$, and work done against overburden pressure in voiding, $U_V$, so that
\begin{equation}
V(w) = U_B + U_V \equiv \frac{B}{2}\int^{\infty}_{-\infty} \frac{w_{xx}^2}{(1 + w_x^2)^{5/2}} dx + q\int^{\infty}_{-\infty}(w-f)\,dx,
\label{eqn:Energy1}
\end{equation} 
where $B$ is the elastic bending stiffness of the layer and $q$ is the overburden pressure per unit length. Note that $U_B$ and $U_V$ are measured from different zero-energy configurations; the former is zero in the flat state and the latter zero in the V-shape. For finite length beams, the two can be reconciled by noting that the work done {\em by} $q$ in moving from the flat beam configuration to the displaced state is a constant (the area of the triangle between the V and the flat beam) {\em minus} the work done {\em against} $q$ in voiding: constants of energy are known not to affect equilibrium equations. The same idea extends in the limit to beams of infinite length. 

By considering stationary solutions of~\eqref{eqn:Energy1} with respect to vertical displacement, equilbrium for the system is described by the Euler-Lagrange equation, 
\begin{equation}\label{EL}
B\left[\frac{w_{xxxx}}{(1+w_x^2)^{5/2}} - 10\frac{w_xw_{xx}w_{xxx}}{(1+w_x^2)^{7/2}} - \frac{5}{2}\frac{w_{xx}^3}{(1+w_x^2)^{7/2}}
+ \frac{35}{2}\frac{w_{xx}^3w_x^2}{(1+w_x^2)^{9/2}}\right]+ q = \mu,
\end{equation}
where $\mu$ is a Lagrange multiplier, arising from the point-wise obstacle constraint $w \geq f$. Physically this quantity can be interpreted as the vertical component of the reactive contact load with the boundary $f$; therefore $\mu$ is zero wherever there is no contact, it matches the overburden pressure $q$ along stretches of contact, and equals the discontinuous shear force at points of delamination. Dodwell {\em et al.}\ \cite{voids1} prove that there exists a unique minimising solution to the Euler-Lagrange equation (\ref{EL}), and that such solutions are convex and symmetric, and they have a single finite interval in which they are not in contact with the boundary. 

Equation \eqref{EL} is written above in Cartesian coordinates $(x,w)$, but exists also in intrinsic coordinates $(s,\psi)$, where $s$ is arc length and $\psi = {\rm d}w/{\rm d}s$. Both representations have their uses: intrinsic coordinates give simpler equations, but for multi-layered problems they carry the added complication of a different coordinate frame for each layer.

\begin{figure}[ht] 
   \centering
  \includegraphics[width=4in]{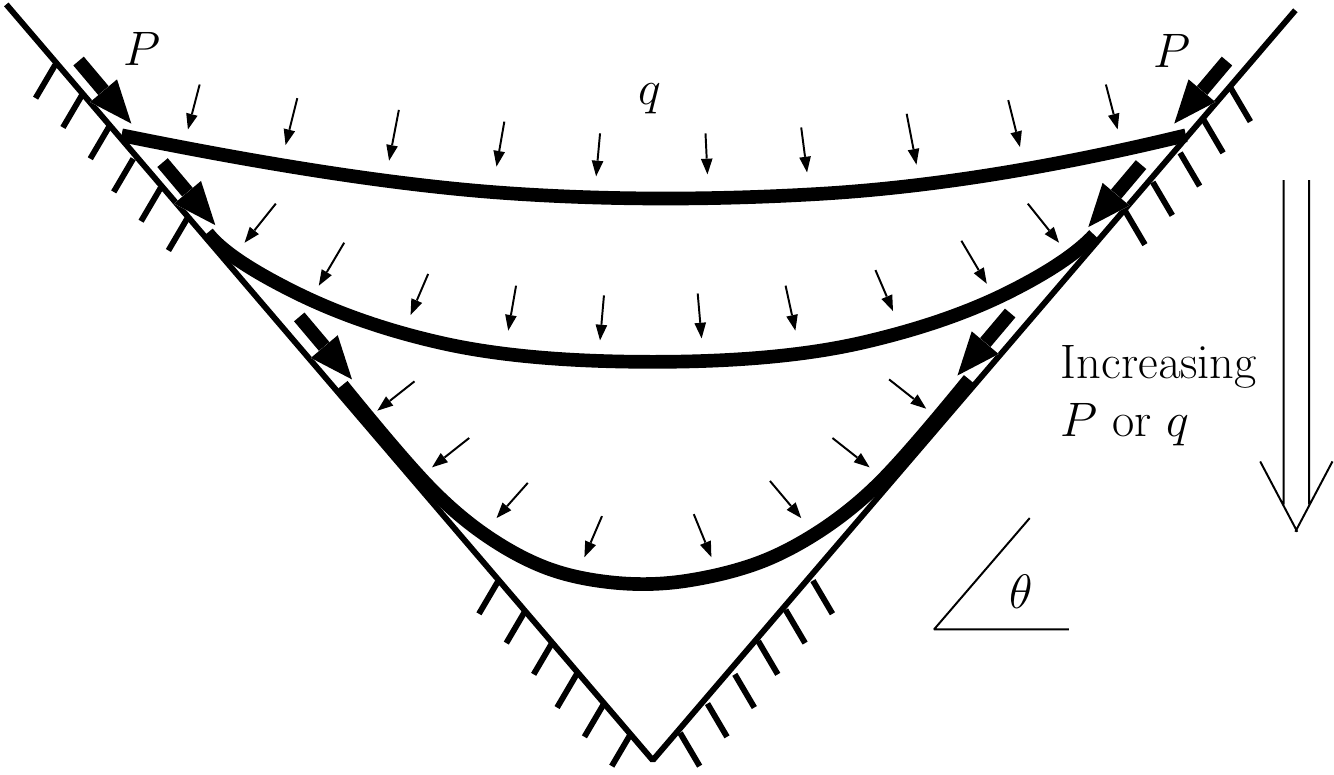}  
     \caption{Finite length beam under pressure loading $q$ and axial compression $P$.}
     \label{fig:configs}
\end{figure}

The formulation of Dodwell {\em et al} \cite{voids1} is readily extended to embrace the different configurations of Fig.~\ref{fig:configs}
for a long beam of finite length $L$, with dead load $P$ acting down the slope. Here we  are primarily interested in the bottom picture, where the combination of $P$ and $q$ has pushed the beam into a similar configuration as Fig.~\ref{fig:modelwithq}, such that some portion of the length is in contact with the supporting medium. The shortening $\Delta$, compared to the flat state $w_x=0$, and measured along the limb,  is given by
\begin{equation}
\label{def:shortening}
\Delta =  - \ell\sqrt{1 + k^2} + \frac{L}{2}\bigl(\sqrt{1+k^2}-1\bigr)  + \int_{0}^{\ell}\sqrt{1 + w_x^2}\;{\rm d}x,
\end{equation}
where $\ell$ is the horizontal non-contact length as seen in Fig.~\ref{fig:modelwithq}. The work done by the compressive load $P$ is then $P\Delta$, and after ignoring a constant the total potential energy over the non-contact region can be written,
\begin{eqnarray}
V(w) = U_B + U_V - 2 P\Delta  \equiv \frac{B}{2}\int^{\ell/2}_{-\ell/2} \frac{w_{xx}^2}{(1 + w_x^2)^{5/2}} dx + q\int^{\ell/2}_{-\ell/2}(w-f)\,dx\nonumber \\
+ \,2\,P\left( \ell\sqrt{1 + k^2} - \int_{0}^{\ell}\sqrt{1 + w_x^2}\;{\rm d}x\right).
\label{eqn:Energy2}
\end{eqnarray} 
This has the Euler-Lagrange equation,
\begin{multline}\label{eqn:ELload}
B\left[\frac{w_{xxxx}}{(1+w_x^2)^{5/2}} - 10\frac{w_xw_{xx}w_{xxx}}{(1+w_x^2)^{7/2}} - \frac{5}{2}\frac{w_{xx}^3}{(1+w_x^2)^{7/2}} + \frac{35}{2}\frac{w_{xx}^3w_x^2}{(1+w_x^2)^{9/2}}\right] +\\+P\left[\frac{w_{xx}}{(1 + w_x^2)^{1/2}} - \frac{w_x^2w_{xx}}{(1 +w_x^2)^{3/2}}\right]+ q = 0.
\end{multline}
Note that the effect of the dead load $P$ can be replaced by rigid load more closely resembling the experimental set-up, by introducing the extra constraint $\Delta = $ constant.
\subsection{Numerical solutions}\label{subsec:numerics}

We now consider solutions which minimise $V$ subject to the constraint $\Delta =$ constant. With the additional loading constraint the favourable properties of minimisers  for the purely pressured case (unconstrained minimization) may cease to hold: constrained minimizers of $V$ are not necessarily unique, symmetric, or convex. Indeed, although there remains the possibility of higher order, asymmetric solutions~\cite{Peletier2001,Champneys1993}, we shall here restrict ourselves to symmetric solutions that have a single finite interval for which they are not in contact with the obstacle $f$.

The Lagrange multipler $\mu$ can be eliminated by writing equation~\eqref{eqn:ELload} as a free-boundary problem~\cite{voids1}. 
Numerically this problem can then be treated as an initial boundary problem with at most two unknown \emph{shooting parameters}, $\ell$ and $P$, depending on the form of loading. Solutions of~\eqref{eqn:ELload} are found by shooting from the  free boundary at $x=-\ell$ with the initial conditions
\begin{equation}
\label{bc:free}
w(-\ell) = k\ell,\quad w_x(-\ell) = k,\quad w_{xx}(-\ell) = 0, \quad w_{xxx}(-\ell) = -(1 +k^2)^{5/2}\frac 1B\left[P\frac{k}{\sqrt{1+k^2}} +  q\ell\right],
\end{equation}
towards the target  symmetric section at $x=0$ given by
\begin{equation}
\label{bc:fixed}
w_x(0) = 0
\qquad \text{and} \qquad w_{xxx}(0) = 0.
\end{equation}
This initial conditions~\eqref{bc:free} can be found by integrating~\eqref{eqn:ELload} once, and imposing continuity of $w$, $w_x$, and $w_{xx}$ at points of delamination $x = \pm\ell$.
\begin{figure}[!h] 
\centering
\quad
  \includegraphics[width=5in]{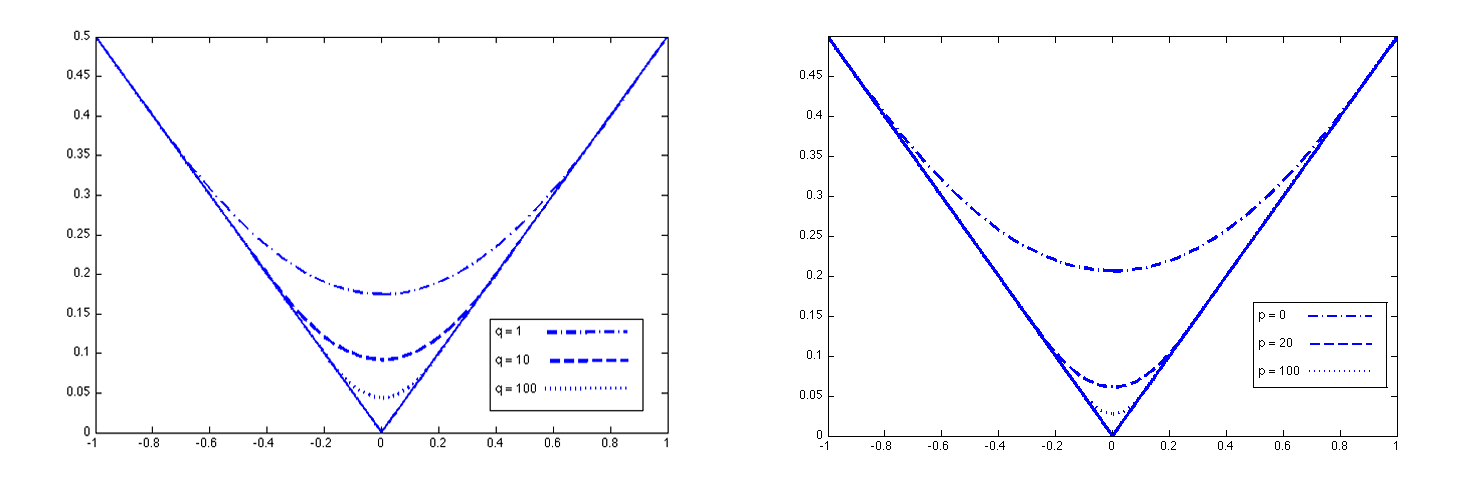} 
     \caption{Equilibrium configurations for $k=1$. (a) $P=0$ and $0 < q < 100$. (b) Primary solutions for $q=1$ and $0 < P < 50$. }     \label{fig:qandPloading}
\end{figure}

Fig.~\ref{fig:qandPloading} shows a number of numerically obtained solutions of this free boundary problem. Fig.~\ref{fig:qandPloading}(a) gives the unique equilibrium solutions for $P=0$ and different values of $q$, whereas Fig.~\ref{fig:qandPloading}(b) displays plots of one of the equilibrium solutions for $q=1$ and increasing $P$. 

\begin{figure}[!h]
  \centering
  \psfig{figure=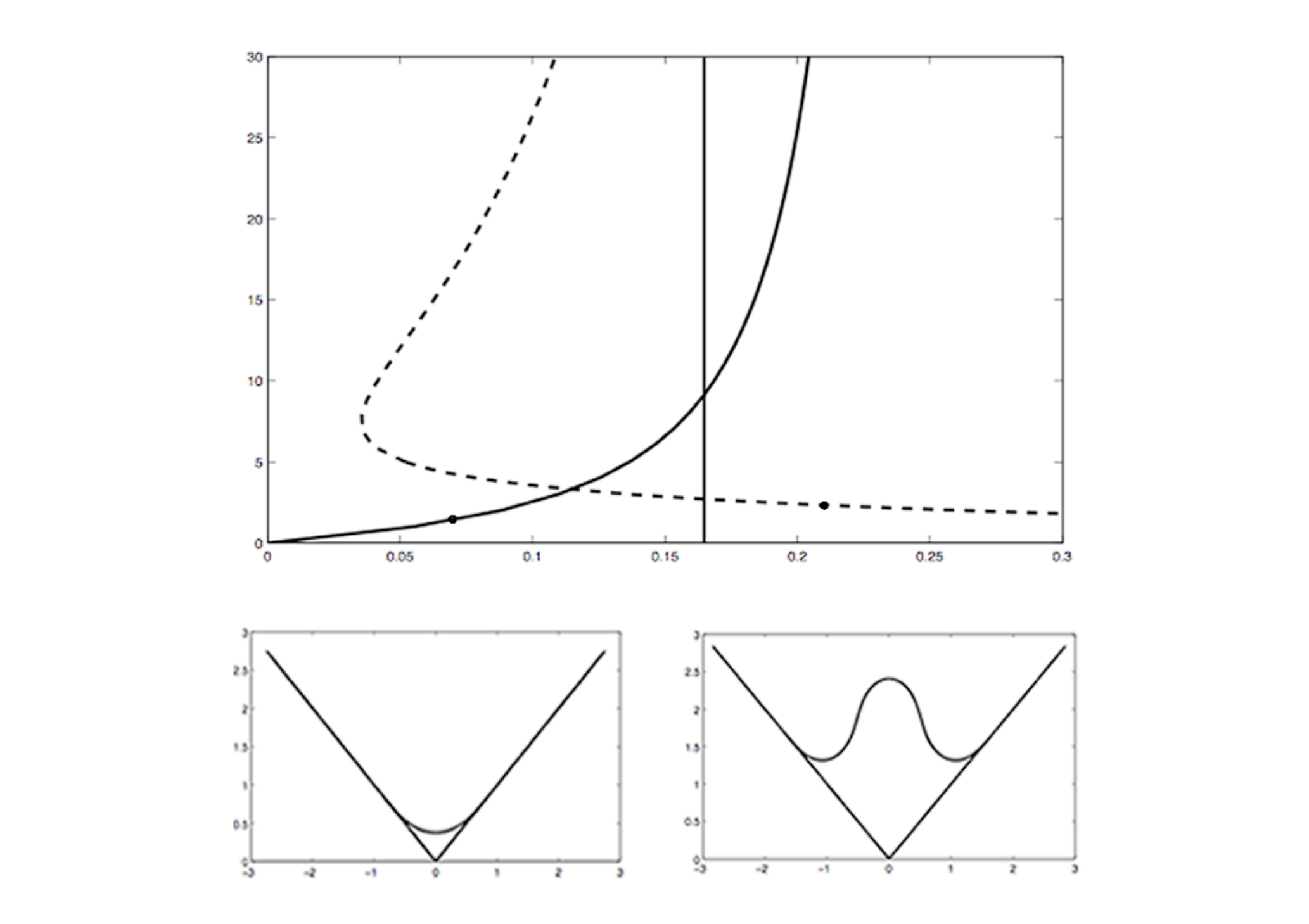,width=6in} 
\caption{The main graph shows values of $P$ for given $\Delta$ for two types of solutions of the Euler-Lagrange equation~\eqref{eqn:ELload}, for fixed pressure $q = 1$ and angle of limb $f_x = k = 1$. The smaller plots  show the profile of the layer at the positions indicated. $\Delta_M$ marks the position of a {\em Maxwell displacement}, where energy levels in the two post-buckled equilibrium states are the same \cite{kinks3}: for $\Delta >  \Delta_M$, minimum energy rests with the upbuckled shape. 
}
\label{fig:loaddisplacement}
\end{figure}

\begin{figure}[!h]
	\centering
		\includegraphics[height=2.5in]{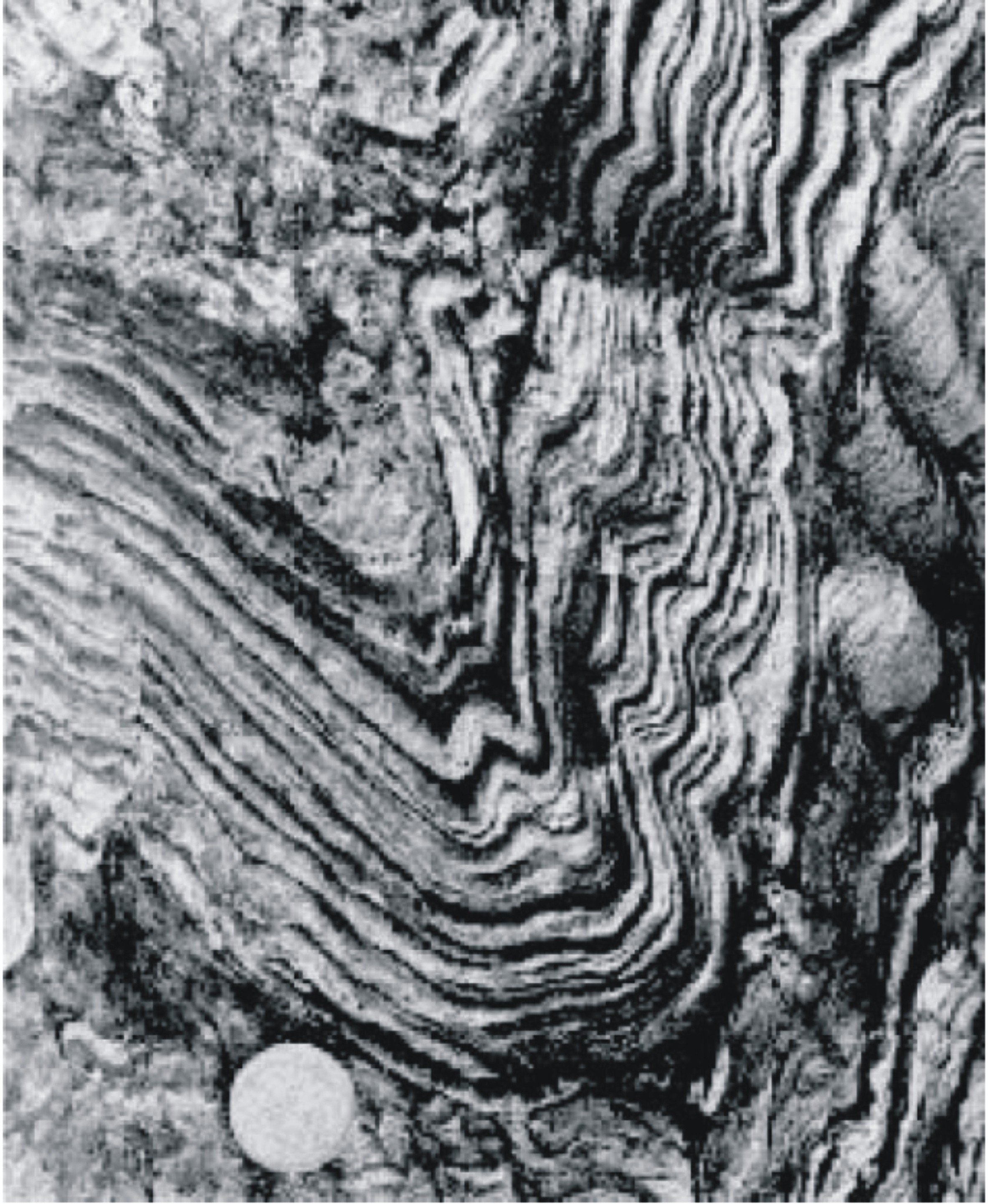}
	\caption{The formation of higher-order accommodation structures in the hinge regions of a multilayered buckle. Rock sample from N. Wales, after Price \& Cosgrove  \protect\cite{PriceCosgrove90}.}
	\label{fig:paraexample}
\end{figure}

Fig.~\ref{fig:loaddisplacement} shows numerical solutions of~\eqref{eqn:ELload}, on a plot of load $P$ against its corresponding deflection, as measured by $ \Delta$, at a constant value of $q$. This shows two of the many possible types of solution. From $P=0$, we first follow the solid line corresponding to the simple convex solution of \ref{fig:qandPloading}(b). As the layer is forced into the corner singularity we observe that  $\Delta\rightarrow L/2$ and $P\rightarrow\infty$. The vertical line in Fig.~\ref{fig:loaddisplacement} marks the Maxwell displacement, $\Delta_M$, where energy levels in the two equilibrium states are the same~\cite{kinks3}. If $\Delta > \Delta_M$, minimum energy rests with the upbuckled shape. In particular, $\Delta = L/2$ gives a upper bound for the existence of simple convex solutions beyond which only higher-order solutions, such as that indicated by the line-dash curve, exist. It is worth noting that such higher modes apparently do exist in geological structures, known as \emph{accommodation structures}~\cite[p.~381]{PriceCosgrove90}. Fig.~\ref{fig:paraexample}
shows one example from a rock outcrop in North Wales. The possibility of energy minimisers involving buckling on a smaller scale is particularly interesting, but we leave the investigation of such convoluted solutions to a further contribution~\cite{parasitic1}.

\section{Multilayered bending and buckling}\label{sec:multi}

\subsection{Multilayer formulation}
\begin{figure}[!h]\quad
	\centering
	\psfig{figure=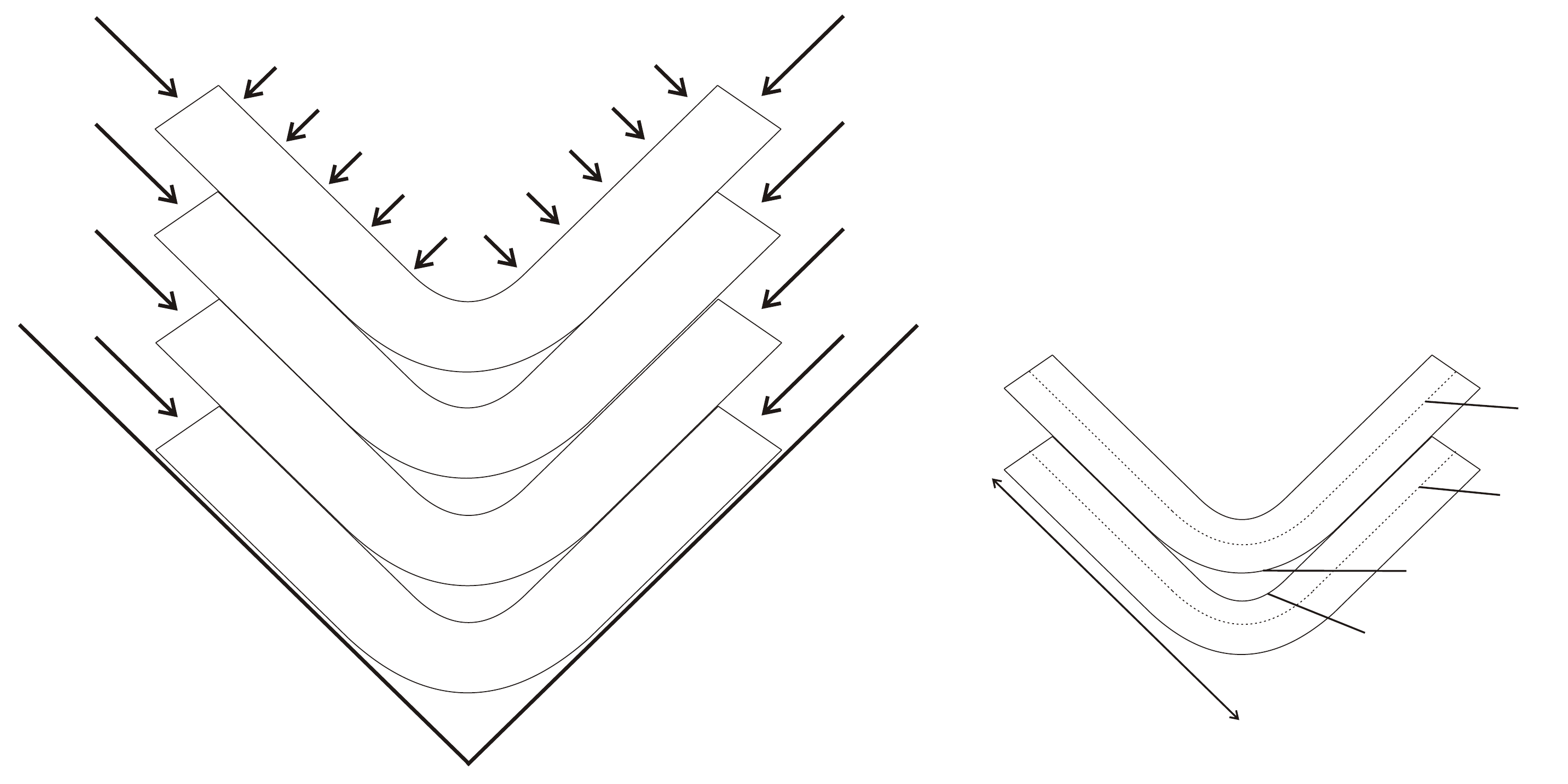,width=6in}
	\caption{Multilayered setup. Here $N$ layers are forced into a V-shaped boundary by an external horizontal load $P$ and pressure $q$.}
	\label{fig:multilayereddeformation}
\end{figure}

We now extend the single-layered model to a multilayered formulation. Consider a stack of $N$ identical layers of uniform thickness $h$ and length $L$ forced into a V-shaped singularity by an overburden pressure gradient defined by $q$ on the top layer, and axial load $P$ per layer, as seen in Fig.~\ref{fig:multilayereddeformation}. 

The centreline of the $i^{th}$ layer from the bottom is labelled as $w^i(x)$, and the top and bottom boundary of that layer as $w^i_t(x)$ and $w^i_b(x)$ respectively. A constant axial load $P$ acts on the stacks, which shortens the $i^{th}$ layer by a distance parameterized by $\Delta_i$.  As before, $\Delta_i$ is measured from the centre of the fold, and here additionally $\xi_i$ gives the horizontal length of the layer. We also introduce the obstacle constraint $w^i_t (x)\leq w^{i+1}_b(x)$ for $i = 1$ to $N-1$ and $w^1(x) \geq f$, so that adjacent layers cannot interpenetrate or pass through the boundary $f = k|x|$. 

We now construct a potential energy function for the system, by adding together the contributing parts.

\subsection{Bending energy}
The total bending energy of the layer is the sum of bending energy of each layer, given by
\[
U_B =  \frac B2 \sum_{i=1}^N \int_{\xi_i}\frac{(w_{xx}^i)^2}{(1 + (w_x^i)^2)^{5/2}}\,dx. 
\]
The quadratic dependence on $w_{xx}^i$ implies that sharp corners have infinite bending energy, and therefore for finite overburden pressure, layers will void rather produce singular corners.

\subsection{Work done against overburden pressure}

In experiments with paper, where voids occur there is clearly a subtle interplay between release in overburden pressure in these areas, and some compensatory increase in internal strain energy elsewhere. Rather than attempt to model this complex situation directly, we simply assume here that the creation of all voids does work against a constant pressure $q$, regardless of position in the sample. Further development is clearly necessary for application to experiments like that of Fig.~\ref{fig:test1-3}, but we leave this for future work. 

The overburden pressure acting on the layers is $q$ per unit length, and therefore the total work done against overburden pressure is $q$ multiplied by the sum of all voids between the layers, given by
\[
U_v = q\int_{\xi_1} (w^1 - f)\;dx + q\sum_{i = 1}^{N-1}\int_{\xi_i}(w^{i+1}_b - w^i_t)\, dx.
\]
The first term gives the energy associated with the void formed between the bottom layer $w^1$ and the boundary $f$, and the second calculates the size of the voids between each of the remaining layers. If $q$ is large, $U_v$ implies a severe energy penalty.

The fundamental assumption of engineering bending theory, that plane sections remain plane and normal to the centreline, means the top and bottom boundaries of a layer can be calculated parametrically by propagating the centerline forwards and backwards a distance $h/2$. In this case it is necessary for the bottom of one layer, $w^{i+1}_b$, and the top of the next, $w^i_t$, to be defined according to the common fixed coordinate system $(x,w)$. At a given $x$ we wish to calculate the void between the $i^{th}$ and $(i+1)^{th}$ layer, defined as $v^i(x) =w^{i+1}_b - w^i_t$.
\begin{figure}[!h]
	\centering
		 \psfig{figure=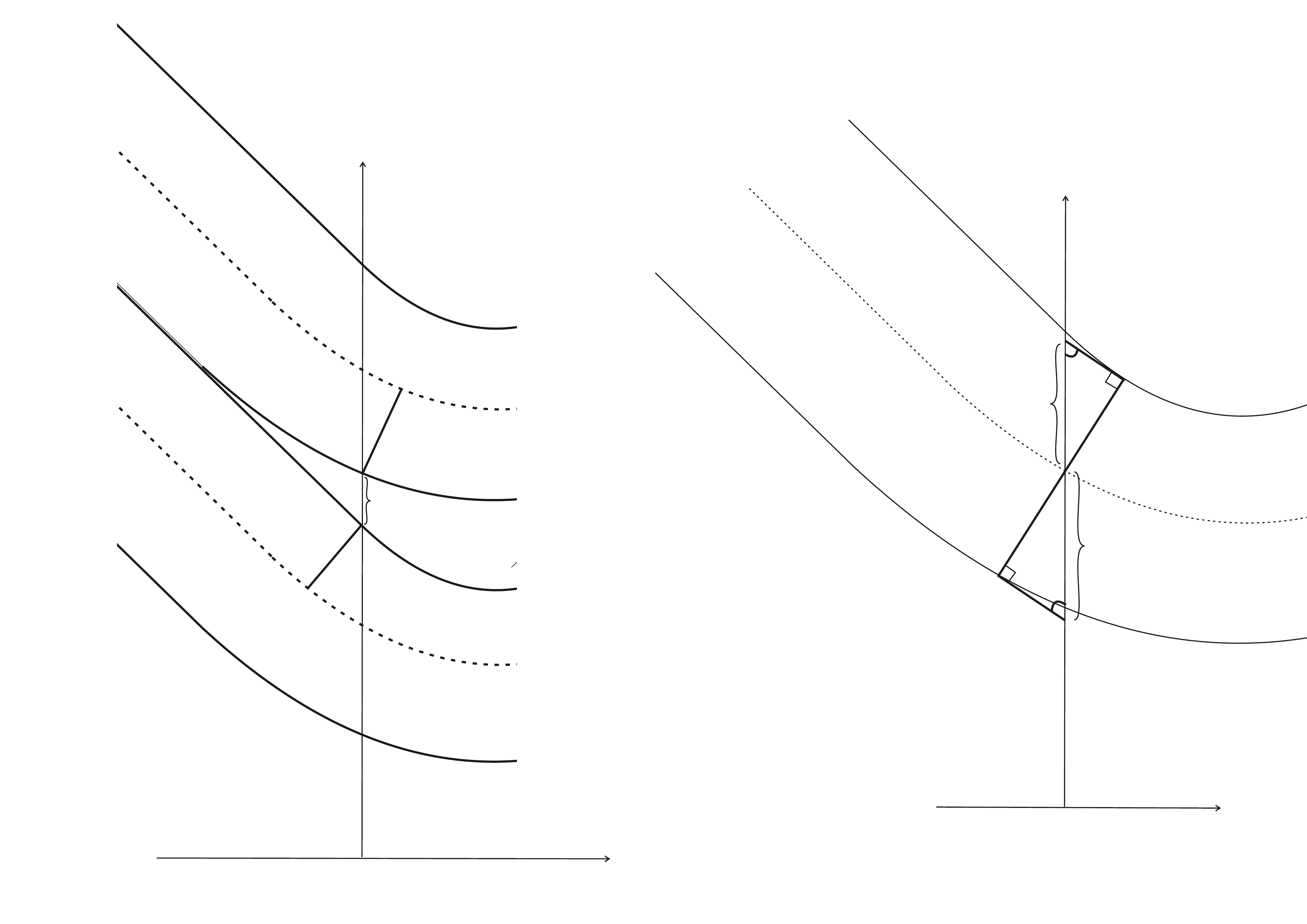,width=5in}  
		\caption{a) Boundary functions determined via front propagation. b) First-order approximation.}
	\label{fig:layerapprox} \quad	
\end{figure}
Figure~\ref{fig:layerapprox} (a) shows that $w^{i+1}_b$ can be determined by propagating $w^{i+1}$ backwards from the point $i)$, and $w^i_t(x)$ by propagating $w^i$ forwards from $ii)$. These centreline positions are not known a priori, but can be reverse-engineered from the common position $x$ by the first-order approximation
\begin{equation}\label{eqn:approx}
w^i_b \approx w^i - \frac h2 \sqrt{1 + (w^i_x)^2} \qquad w^i_t  \approx w^i + \frac h2 \sqrt{1 + (w^i_x)^2}.
\end{equation}
Fig.~\ref{fig:layerapprox} (b) demonstrates this approximation, where $iii)$ and $iv)$ are found by constructing  right-angle triangles. This is a thin-layer approximation, in which we assume that the layer is much thinner than the radius of curvature of the layer. Whether that is justified or not depends on the situation. We also note that the approximation is exact for straight segments, whereas for the convex segment shown in Fig.~\ref{fig:layerapprox} b), $iii)$ is an underestimate and $iv)$ an overestimate: and vice versa for a concave segment. With this approximation the total energy in voiding becomes
\[
U_v =  q\int_{\xi_1} (w^1 - f)\;dx + q\sum_{i = 1}^{N-1}\int_{\xi_i}\Bigl[w^{i+1} -\frac h2
\sqrt{1 + (w^{i+1}_x)^2} - w^i - \frac h2 \sqrt{1 + (w^i_x)^2}\Bigr]\, dx.
\]

\subsection{Work done by load}
If we consider a constant axial load $P$ acting on each layer of the multilayered stack, the total work done by the load is the sum of the work done by the load in the shortening of each layer. Therefore up to a constant the total work done by the load is,
\[
U_P = 2P\sum_{i=1}^N\Delta_i = -P\sqrt{1 + k^2}\sum _{i=1}^N \xi_i.
\]

\subsection{Total potential energy function}
The total potential energy for the complete system can be written as the sum of work done in bending by each layer, total work done against overburden pressure in voiding, and work done on the system by the axial load $P$ in shortening, as follows:
\begin{align}\label{eqn:potentialenergyMultilayer} 
\nonumber 
V(w_1,\ldots,w_n) &= U_B + U_V +U_P \\ \nonumber
&=\frac B2 \sum_{i=1}^N \int_{\xi_i}\frac{(w_{xx}^i)^2}{(1 + (w_x^i)^2)^{5/2}}\, dx +q\int_{\xi_1} (w^1 - f)\;dx\\
 &+ q\sum_{i = 1}^{N-1}\int_{\xi_i}(w^{i+1}_b - w^i_t)\, dx - 2\sum_{i=1}^NP\Delta_i.
\end{align}
Appropriate differentiation leads to a system of $N$ nonlinear fourth order equations, coupled by contact conditions. 

\section{Periodic solutions}\label{sec:periodic}

The formulation given above gives a general energy description for $N$ identical layers forced into a V-shaped singularity. Now, motivated by Fig.~\ref{fig:test5}, we restrict ourselves to minimizing $V$ over the class of symmetric periodic functions with one void per period. We also make the assumption that layers are symmetric, with straight limbs of fixed angle $k$ connected by a convex curved section. From Section~\ref{subsec:numerics} we recognise that higher-order solutions and  asymmetric solutions might exist, but the experiments shown in Fig.~\ref{fig:test5} suggest that the assumption of this simplified geometry is justified at least in some limited context.

\subsection{$1$-periodic solutions}
\begin{figure}[H]
\quad
\centering
 \psfig{figure=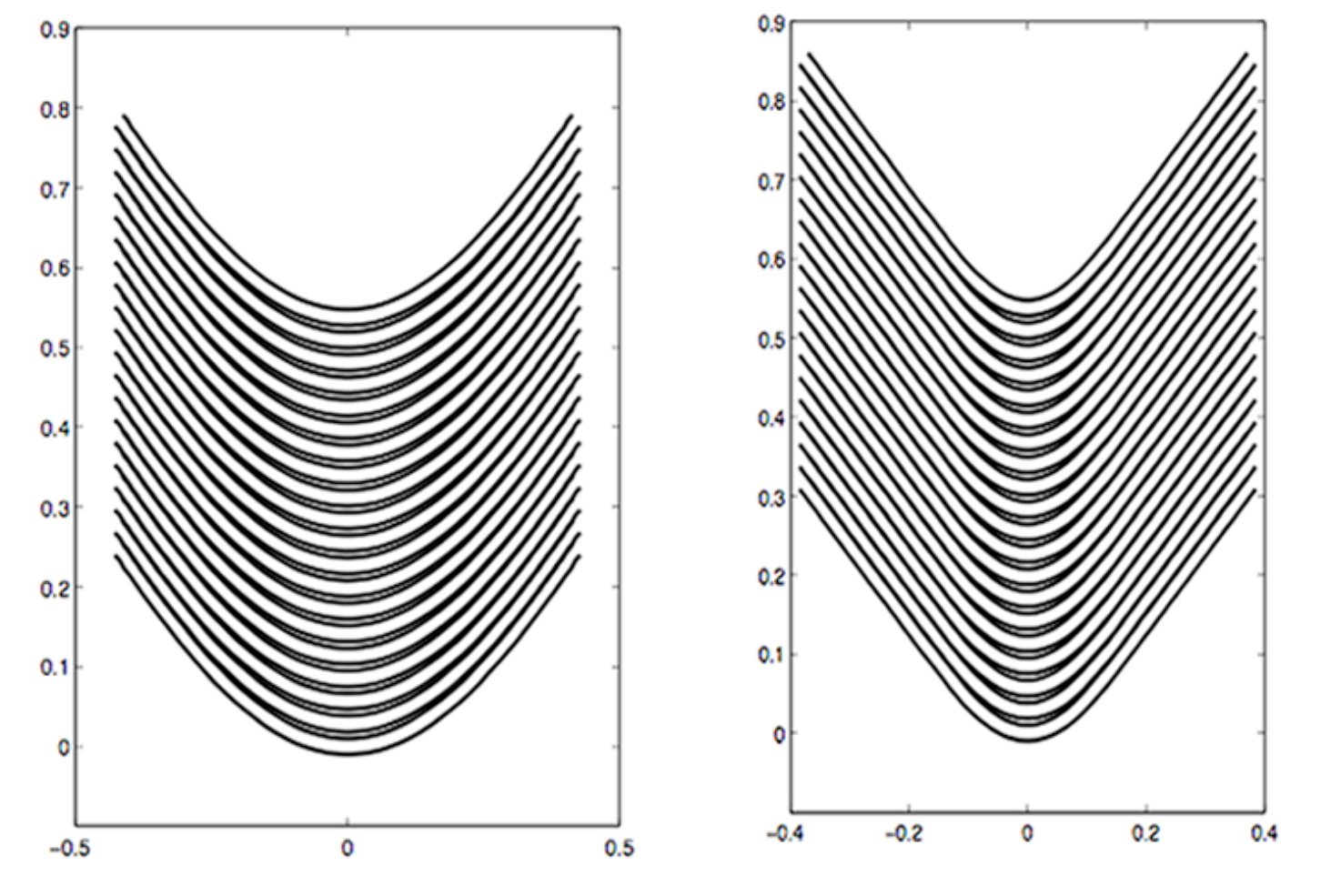,width=5in}  
\caption{Two examples of solution profiles for fixed angle $k = 0.75$, $h= 0.02$, load $P=1$, and two values of overburden pressure; (left) $q = 10$ and (right) $q = 100$.}
\label{fig:solution}\quad
\end{figure}

We start with the simplest periodic solution where the structure repeats every layer, hereafter named a $1$-periodic solution. Since each layer deforms identically the total potential energy may be described in terms of a single layer. Each layer does the same work in bending and shortening, and the size of the void can be calculated by shifting the bottom boundary of a layer to fit above the top of the same layer. Thus, for the primary buckling mode shape of Fig.~\ref{fig:qandPloading}(b),
 \[
 w^{i+1}_b = w^i_b   + h \sqrt{1 + k^2}.
 \]
The total potential energy~\eqref{eqn:potentialenergyMultilayer} simplifies to
  \begin{align}\label{eqn:singleperiodenergy1}
V(w) &= \frac {BN}2\int_{\xi} \frac{w_{xx}^2}{(1 + w_x^2)^{5/2}}\,dx +q\int_{\xi} (w - f)\,dx-q(N-1)h\int_{\xi}\sqrt{1 + (w_x)^2}\,dx - NP\Delta,
\end{align}
where the shortening $\Delta$ is again given by~\eqref{def:shortening}.
Minimisers of~\eqref{eqn:singleperiodenergy1} solve the Euler-Lagrange equation over the non-contact region, given by
\begin{multline}\label{eqn:ELperiodic}
BN\left[\frac{w_{xxxx}}{(1+w_x^2)^{\frac{5}{2}}} - 10\frac{w_xw_{xx}w_{xxx}}{(1+w_x^2)^{\frac{7}{2}}} - \frac{5}{2}\frac{w_{xx}^3}{(1+w_x^2)^{\frac{7}{2}}} + \frac{35}{2}\frac{w_{xx}^3w_x^2}{(1+w_x^2)^{\frac{9}{2}}}\right]  \\+ (q(N-1)h + 2P)\left[\frac{w_{xx}}{(1 + w_x^2)^{1/2}} - \frac{w_x^2w_{xx}}{(1 +w_x^2)^{3/2}}\right] + q= 0.
\end{multline}
As in Section~\ref{sec:singlemodel}, boundary conditions can be found by integrating~\eqref{eqn:ELperiodic} with respect to the spatial coordinate $x$ and imposing continuity of $w_x$ and $w_{xx}$ at points of delamination $ x= \pm\ell$, so that
\begin{equation}
\label{bc:multifree}
w(-\ell) = k\ell,\quad w_x(-\ell) = k,\quad w_{xx}(-\ell) = 0
\end{equation}
and
\[
w_{xxx}(-\ell) = -(1 +k^2)^{5/2}\frac 1{BN}\left[(qh(N-1) + 2NP)\frac{k}{\sqrt{1+k^2}} +  q\ell\right],
\]
coupled with the symmetric boundary condition at $x=0$. Figure~\ref{fig:solution} shows examples of solution profiles obtain by numerical  shooting methods. The results match physical intuition in the sense that as overburden pressure $q$ or load $P$ increases the void size decreases, leaving in the limit deformations with straight limbs and sharp corners.

\subsection{$m$-periodic solutions}
The previous section naturally extends to $m$-periodic solutions, where a void forms after $m$ layers and then the structure repeats. This leads to various interesting questions. Is there a value of $m$, not always $1$ or $\infty$, which has minimium energy among the class of such periodic solutions? Can we predict the value of $m=125$ seen in the periodic solutions presented in Fig~\ref{fig:test5}? As we have seen, $1$-periodic solutions can be solved using previously developed analysis~\cite{voids1}. This is not the case for more general $m$-periodic solutions, where deriving the energy and the subsequent variational analysis is a non-trivial exercise. We therefore leave the details of this work to a future publication. To test the basic ideas, however,  we perform a  very rough calculation for the experiment presented in Fig.~\ref{fig:test5}, to see if a simple energy analysis, with reasonable values of the physical constants, gives a value of $m$ of the correct order of magnitude. 

\begin{figure}[h]
\quad
\centering
 \psfig{figure=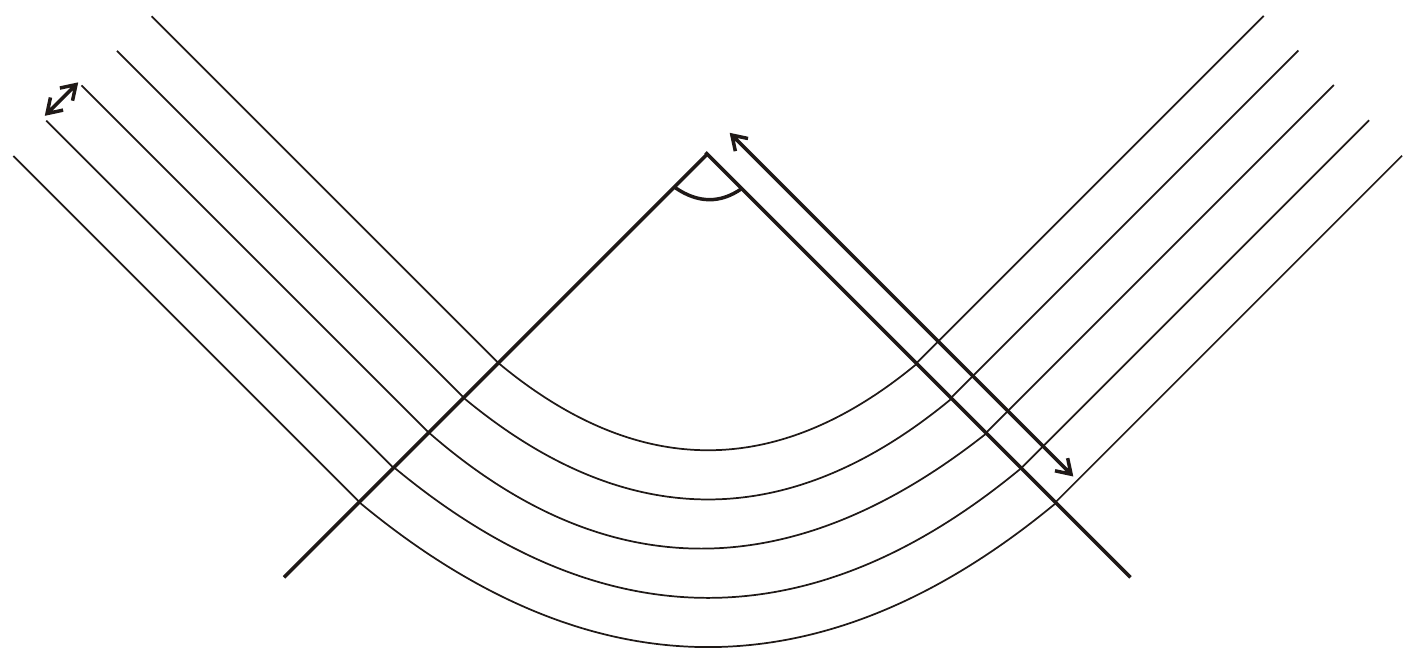,width=5in}  
\caption{Shows the construction of an $m$-periodic package as straight limbs connected by circular arcs. Here the thickness of each layer is $h$ and the radius of curvature of the out side arc is given by $R$.}
\label{fig:mperiodsetup}
\end{figure}

The central premise is that in equilibrium the bending energy in a single $m$-periodic packet is approximately equal to the energy required to form one of the voids. For this analysis we assume that each layer has straight limbs connected by a circular arc (see Fig.~\ref{fig:mperiodsetup}) 
where the circular arc of the outside layer is an arc of radius $R$ swept through an angle $2\theta$. The angle $\theta$ is fixed for given gradient of the limbs $k$. The bending energy stored in an $m$-periodic packet of layers is therefore
\begin{equation}\nonumber
U_B = \frac{EI}{2}\sum^{m}_{i=1}\int^L_0\kappa_i(s)^2\;ds_i = \frac{EI}{2}\sum^{m}_{i=1}\int^{\theta}_{-\theta}\kappa_i(\theta)\;d\theta,
\end{equation}
where $\kappa_i$ is the curvature  and $s_i$ is the arc length of the $i^{th}$ layer. Here we number the layers as $1$ to $m$, from the outside inwards. Following a result presented in~\cite{levelsets},  $\kappa_i = \kappa_1/(1 - (i-1)h\kappa_1)$, and it therefore follows that
\begin{equation}\nonumber
U_B = \frac{EI}{2}\sum^m_{i+1}\int^{\theta}_{-\theta}\frac{\kappa_1}{(1 - (i-1)h\kappa_1)}\;d\theta = \frac{EI}{2}\int^{\theta}_{-\theta}\frac{1}{h}\underbrace{\sum^m_{i=1}\frac{h\kappa_1}{1 - h(i-1)\kappa_1}}_{:=f(\kappa_1)}\;d\theta.
\end{equation}
For large $m$, we  consider $f(\kappa_1)$ to be a Riemann sum, and approximate it by the corresponding integral, i.e. setting $x_i = h(i-1)$,
\[
\text{we approximate}\quad 
\sum^m_{i=1}\frac{h\kappa_1}{1 - x_i\kappa_1}
\quad\text{by}\quad
\int_{x_1}^{x_{m+1}} \frac{\kappa_1}{1 - x\kappa_1}\, dx 
= - \log (1-x\kappa_1)\Big|_{x=x_1}^{x=x_{m+1}},
\]
by which
\begin{equation}\nonumber
U_B = -\frac{EI}{2h}\int^{\theta}_{-\theta} \log(1 - mh\kappa_1)\;d\theta.
\end{equation}
For a circular arc $\kappa_1 = 1/R$ is constant over $\theta$, therefore
\begin{equation}\label{eqn:mperiodbending}
U_B = -\frac{EI}{h}\log\left(1-\frac{mh}{R}\right)\theta.
\end{equation}
Estimating the work done to produce a void is simply the overburden pressure $q$ times the size of the void, we can equate the forms of energy, so that
\begin{equation}\label{eqn:energybalance}
-\frac{EI}{h}\log\left(1-\frac{mh}{R}\right)\theta \approx q\times\mbox{Void Area}
\end{equation}
The table below displays estimates of the physical constants for the experiment, and gives a brief description of their source.
\begin{center}
\begin{tabular}{c|c|c}\label{tab:parameters}\centering \textbf{Constants} & \textbf{Values} &  \textbf{Source} \\ \hline E & $5$kNmm$^{-2}$ & \cite{kinks3} \\ I & $h^3/12$ & Standard Calculation \\Transverse Load & $40kN$ & Fig. \ref{fig:kinks}. \\Area& $3\times10^4$mm$^2$ & Area of A5 paper  \\q& $1.33$Nmm$^{-2}$ & Transverse Load/Area \\Void Area & $10$mm$^{2}$ &  Fig.~\ref{fig:test5}. \\$\theta$ & $ 1$ & Fig.~\ref{fig:test5}. \\ Radius of Curvature & $5$mm & Fig.~\ref{fig:test5}. \\ $h$  & $0.1$mm  & Thickness of A5 paper  
\end{tabular}
\end{center}
Subsituting in the parameter from the table above into~\eqref{eqn:energybalance} we obtain
\[
\log\left(1-\frac{mh}{R}\right) \approx -2,
\]
giving a rough estimate of $m \approx 40$. 

The fact that an intermediate value of $m$ is chosen suggsts that the basic premise of the calulation (balance of bending and voiding energy) might be reasonable. The fact that the predicted number $m\approx 40$ is in the same ballpark as the observed value of $125$ in Fig.~\ref{fig:test5} further reinforces this suggestion.

\section{Concluding Remarks}

This paper extends a single-layered model for folding and void formation \cite{voids1} to the simplest periodic multilayered structure, including the vital effects of axial loading. The model extends our insight into the interplay between material and loading properties on one hand and the geometric constraints of layers fitting together on the other. The results match physical intuition, in that as overburden pressure increases, void size decreases,  in the limit leading to deformations with straight limbs and sharp corners. Although an  oversimplification in terms of  the paper experiments, and even more so of their true geological counterparts, it does capture many of their important hallmarks, as well as demonstrating the possibility of periodic solutions arising naturally in an elastic context. 

Studying these $1$-periodic solutions has led to  interesting questions about the possibility of $m$-periodic solutions. In particular, can we find which $m$-periodic solution has minimum energy? The simple energy argument presented in the previous section gives a  rough estimate for $m$ for the experiment presented in Fig.~\ref{fig:test5}. This gives an interesting path of research to follow, and one we wish to consider in a future contribution.

\bibliography{refs}

\begin{thebibliography}{10}

\bibitem{BeexPeer}
L.~A.~A. Beex and R.~H.~J. Peerlings.
\newblock On the influence of delamination on laminated paperboard creasing and
  folding.
\newblock {\em Phil.\ {T}rans.\ {R}. {S}oc.\ {L}ond.}, 2011.
\newblock This Special Issue.

\bibitem{levelsets}
J.~A. Boon, C.~J. Budd, and G.~W. Hunt.
\newblock Level set methods for the displacement of layered materials.
\newblock {\em Proc.\ {R}. {S}oc.\ {L}ond.}, {\rm A} 463:1447--1466, 2007.

\bibitem{parallel}
C.~J. Budd, R.~Edmunds, and G.~W. Hunt.
\newblock A nonlinear model for parallel folding with friction.
\newblock {\em Proc.\ {R}. {S}oc.\ {L}ond.}, {\rm A} 459:2097--2119, 2003.

\bibitem{Champneys1993}
A~R Champney and J~F Toland.
\newblock Bifurcation of a plethora of multi-modal homoclinic orbits for
  autonomous hamiltonian systems.
\newblock {\em Nonlinearity}, 665(6), 1993.

\bibitem{parasitic1}
T.~J. Dodwell.
\newblock A variational description of convulted accomodation structures in
  geological folds.
\newblock submitted for Philosophical Magazaine: Special Issue: Instability
  across the Scales, 2011.

\bibitem{voids1}
T.~J. Dodwell, M.~A. Peletier, C.~J. Budd, and G.~W. Hunt.
\newblock Self-similar voiding solutions of a bi-layered model of folding
  rocks.
\newblock submitted to SIAM for publication, 2011.

\bibitem{Hobbs2011}
B.~E. Hobbs and A.~Ord.
\newblock Localised and chaotic folding: The role of axial plance structures.
\newblock This Special Issue, 2011.

\bibitem{HudlestonTreagus10}
P.J. Hudleston and S.H. Treagus.
\newblock {Information from folds: A review}.
\newblock {\em Journal of Structural Geology}, 2010.

\bibitem{fold}
G.~W. Hunt, H-B. M{\"{u}}hlhaus, and A.~I.~M. Whiting.
\newblock Evolution of localized folding for a thin elastic layer in a
  softening visco-elastic medium.
\newblock {\em {PAGEOPH (Pure and Appl. Geophysics)}}, 146(2):229--252, 1996.

\bibitem{HMW}
G.~W. Hunt, H-B. M{\"{u}}hlhaus, and A.~I.~M. Whiting.
\newblock Folding processes and solitary waves in structural geology.
\newblock In A.~R. Champneys, G.~W. Hunt, and J.~M.~T. Thompson, editors, {\em
  {Localization and Solitary Waves in Solid Mechanics}}, volume {\rm A} 355,
  pages 2197--2213. Theme Issue of {\em Phil.\ {T}rans.\ {R}. {S}oc.\ {L}ond},
  1997.

\bibitem{HuntPeletierWadee00}
G.~W. Hunt, M.~A. Peletier, and M.~A. Wadee.
\newblock The {M}axwell stability criterion in pseudo-energy models of kink
  banding.
\newblock {\em J.~Structural Geology}, 22:667--679, 2000.

\bibitem{Peletier2001}
M~A Peletier.
\newblock Sequential buckling: A variational analysis.
\newblock {\em SIAM Journal on Mathematical Analysis}, 32:1142--1168, 2001.

\bibitem{Pinho2011}
S.~T. Pinho, R.~Gutkin, S.~Pimenta, N.~V.~De Carvalho, and R.~Robinson.
\newblock On longitudinal compressive failure of cfrp: form unidirectional to
  woven, and from virgin to recycled.
\newblock This Special Issue, 2011.

\bibitem{PriceCosgrove90}
N.~J. Price and J.~W. Cosgrove.
\newblock {\em Analysis of Geological Structures}.
\newblock Cambridge University Press, 1990.

\bibitem{Schmalholz2011}
S.~M. Schmalholz and D.~W. Schmid.
\newblock Folding in power-law viscous multilayers.
\newblock This Special Issue, 2011.

\bibitem{Wadee2011}
M.~A. Wadee, C.~V{\"{o}}llmecke, J.~F. Haley, and S.~Yiatros.
\newblock Geometric modelling of kink banding in laminated structures.
\newblock This Special Issue, 2011.

\bibitem{kinks3}
M.~Ahmer Wadee, G.~W. Hunt, and M.~A. Peletier.
\newblock Kink band instability in layered structures.
\newblock {\em J. {M}ech.\ {P}hys.\ {S}olids}, 52(5):1071--1091, 2004.

\end{thebibliography}

\end{document}